\documentclass[twocolumn,letter]{emulateapj}

\usepackage{graphicx}
\usepackage{latexsym,amssymb}
\usepackage{amsmath}
\usepackage{natbib}
\newcommand{\tw}{Tremaine-Weinberg}

\newcommand{\fantomm}{{\tt FaNTOmM}}
\newcommand{\ghafas}{{\sc GH$\alpha$FaS}}

\newcommand{\CO}{{\sc CO}}
\newcommand{\Ha}{H$\alpha$}
\newcommand{\Op}{$\Omega_p$}

\newcommand{\HI}{{\sc H$\,$i}}
\newcommand{\HII}{{\sc H$\,$ii}}

\newcommand{\Vlos}{$V_\mathrm{los}$}
\newcommand{\Vsys}{$V_\mathrm{sys}$}
\newcommand{\Vrot}{$V_\mathrm{rot}$}
\newcommand{\Vrad}{$V_\mathrm{rad}$}
\newcommand{\pak}{$\phi_k$}
\newcommand{\pap}{$\phi_p$}

\def\kms{$\mbox{km s}^{-1}$}
\def\kmskpc{$\mbox{km s}^{-1}\mbox{ kpc}^{-1}$}
\def\kmsarc{$\mbox{km s}^{-1}\mbox{ arcsec}^{-1}$}
\def\Msun{$\mbox{M}_\odot$}
\def\Myr{$\mbox{M}_\odot\mbox{ yr}^{-1}$}

\slugcomment{Submitted to ApJ May 25, 2009; Accepted \today}

\shorttitle{Pattern Speeds of Bars and Spirals From \Ha\ Velocity Fields} 

\shortauthors{Fathi et al.}

\begin{document}

\title{Pattern Speeds of Bars and Spiral Arms From \Ha\ Velocity Fields $^{\star}$}
\author{Kambiz Fathi\altaffilmark}
\address{Stockholm Observatory, Department of Astronomy, Stockholm University, AlbaNova, 106 91 Stockholm, Sweden}
\address{Instituto de Astrof\'\i sica de Canarias, C/ V\'\i a L\'actea s/n, 38200 La Laguna, Tenerife, Spain}
\address{Oskar Klein Centre for Cosmoparticle Physics, Stockholm University, 106 91 Stockholm, Sweden}

\author{John E. Beckman} 
\address{Instituto de Astrof\'\i sica de Canarias, C/ V\'\i a L\'actea s/n, 38200 La Laguna, Tenerife, Spain}
\address{Consejo Superior de Investigaciones Cient\'\i ficas, Spain}

\author{N\'uria Pi\~nol-Ferrer} 
\address{Stockholm Observatory, Department of Astronomy, Stockholm University, AlbaNova, 106 91 Stockholm, Sweden}
\address{Instituto de Astrof\'\i sica de Canarias, C/ V\'\i a L\'actea s/n, 38200 La Laguna, Tenerife, Spain}

\author{Olivier Hernandez}
\address{LAE, Universit\'e de Montr\'eal, C.P. 6128 succ. centre ville, Montr\'eal, QC, Canada H3C 3J7}

\author{Inma Mart\'\i nez-Valpuesta}
\address{Instituto de Astrof\'\i sica de Canarias, C/ V\'\i a L\'actea s/n, 38200 La Laguna, Tenerife, Spain}

\author{Claude Carignan}
\address{LAE, Universit\'e de Montr\'eal, C.P. 6128 succ. centre ville, Montr\'eal, QC, Canada H3C 3J7}

\altaffiltext{$\star$}{The derived H$\alpha$ rotation curves for all galaxies are made available in electronic format at the CDS via anonymous ftp to {\tt cdsarc.u-strasbg.fr} or via {\tt http://cdsweb.u-strasbg.fr/}.}

\begin{abstract}
We have applied the \tw\ method to 10 late-type barred spiral galaxies using data cubes, in \Ha\ emission, from the \fantomm\ and \ghafas\ Fabry-Perot spectrometers. We have combined the derived bar (and/or spiral) pattern speeds with angular frequency plots to measure the corotation radii for the bars in these galaxies. We base our results on a combination of this method with a morphological analysis designed to estimate the corotation radius to bar-length ratio using two independent techniques on archival near infrared images, and although we are aware of the limitation of the application of the Tremaine-Weinberg method using \Ha\ observations, we find consistently excellent agreement between bar and spiral arm parameters derived using different methods. 
In general, the corotation radius, measured using the \tw\ method, is closely related to the bar length, measured independently from photometry and consistent with previous studies. Our corotation/bar-length ratios and pattern speed values are in good agreement with general results from numerical simulations of bars. In systems with identified secondary bars, we measure higher \Ha\ velocity dispersion in the circumnuclear regions, whereas in all the other galaxies, we detect flat velocity dispersion profiles. In the galaxies where the bar is almost purely stellar, \Ha\ measurements are missing, and the \tw\ method yields the pattern speeds of the spiral arms. 
The excellent agreement between the Tremaine-Weinberg method results and the morphological analysis and bar parameters in numerical simulations, suggests that although the \Ha\ emitting gas does not obey the continuity equation, it can be used to derive the bar pattern speed. In addition, we have analysed the \Ha\ velocity dispersion maps to investigate signatures of secular evolution of the bars in these galaxies. The increased central velocity dispersion in the galaxies with secondary bars suggest that the formation of inner bars or disks may be a necessary step in the formation of bulges in late-type spiral galaxies.
\end{abstract}

   \keywords{ galaxies: spiral -- 
   	      galaxies: kinematics and dynamics -- 
	      galaxies: ISM --
	      galaxies: structure -- 
	      galaxies: evolution --
	      galaxies: individual (IC\,342, NGC\,2403, NGC\,4294, NGC\,4519, NGC\,5371, NGC\,5921, NGC\,5964, NGC\,6946, NGC\,7479, NGC\,7741)}
   \maketitle

\section{Introduction}
The theory of resonant structure in disk galaxies in its original linear form by \citet{Lindblad1963} or in its more complete form by \citet{LinShu1966} implies that a density wave pattern in the stellar disk acts on the rotating gas to form stars continually along spiral shock lines associated with the arms or virtually straight shocks associated with a bar. Additionally, resonances play an important role in the mechanism of maintaining density waves, which are the back-bone of the spiral structure in the inner parts as well as outer regions of spiral galaxies \citep{Lin1970}. One corollary is that over significant ranges of galactocentric radius the angular velocity of the density wave pattern may well be virtually invariant. These invariant patterns give rise to the bar (or each bar if nested bars are present) and to the spiral arms, and their respective angular velocities are termed pattern speeds, \Op. 

Observational derivation of the \Op, together with the conventional rotation curve, allows us to characterise the main dynamical parameters of the disk. \citet{TremaineWeinberg1984TW} suggested a virtually model independent purely kinematic way to derive \Op\, of which a simplified form was given by \citet{MerrifieldKuijken1995}, where the Tremaine-Weinberg method can be written as 
\[\Omega_p \sin(i) =  \langle V\rangle \, / \, \langle X\rangle,\] 
where $i$ is the inclination angle of the disk and the parameters $\langle V\rangle$ and $\langle X\rangle$ are the luminosity-weighted averages of the projected velocity at each pixel, and the projected distance from the minor-axis, respectively, measured along a slit parallel to the line of nodes. The method is based on three assumptions, and only under these assumptions can the relation between $\langle V\rangle$ and $\langle X\rangle$ be used to determine \Op: 
\begin{enumerate}
\item The disk is flat.
\item The disk has a well-defined pattern speed.
\item The surface brightness of the tracer obeys the continuity equation.
\end{enumerate}

To make use of the \tw\ method, we need to take a spectrum along such a slit which gives us the projected velocity, and also to take an image giving the luminosity in the component used to measure the spectrum \citep[e.g., ][]{MerrifieldKuijken1995, Corsini2007}. In practice, obtaining several slit spectra of galaxies is a tedious process, and consequently, two-dimensional velocity fields can be used to simulate a large number of slits, enhancing the signal-to-noise ratio and better exploring the regime of multiple \Op s. To fulfill the third assumption, observations yielding stellar kinematics are considered ideal. However, high resolution observations of the stellar line-of-sight velocities over entire galaxies are still technically challenging \citep[see, e.g., ][]{Emsellem2007}. Alternatively, emission-lines can be used to derive the kinematics of the interstellar medium in a much easier way. However, the interstellar gas is subject to large scale shocks and smaller scale expansions powered by OB associations, and hence does not fulfill the third assumption. 

Several studies have shown that the \tw\ method can be applies to \CO\ velocity fields \citep{RandWallin2004, Zimmeretal2004} or \Ha\ velocity fields \citep{Hernandezetal2005TW, Emsellemetal2006, Fathietal2007TW, CheminHernandez2009, Gabbasov2009} in individual galaxies to derive \Op s in agreement with numerical simulations. Furthermore, these \Op s could be used to identify the location of resonances where predicted morphological features, such as bar-spiral transition or rings, have been identified. The latter studies have suggested that although \Ha\ does not satisfy the third assumption, it may come close enough.

Here we have applied the \tw\ method on Fabry-Perot data cubes in \Ha\ emission which provide, over the full extent of the disks, complete \Ha\ kinematic maps, red continuum maps, as well as continuum-subtracted \Ha\ surface brightness maps for a sample of 10 late-type barred spiral galaxies. Although we are aware of the limitations of such an analysis, the wide spatial coverage combined with the high spatial and spectral resolution makes this data set excellent for this type of analysis. When combined with morphological analysis, this yields an angular speed parameter for the bar which we will show to be a good approach to the pattern speed. Support for this method can be found in \citet[][]{Hernandezetal2005TW} who found a main \Op\ for NGC\,4321 in good agreement with \CO\ studies by \citet{Wadaetal1998} and \citet{RandWallin2004}. Furthermore, \citet{Emsellemetal2006} showed that the derived \Op\ for NGC\,1068 is fully consistent with the stellar velocity field.

In section~\ref{sec:observations}, we describe the observations, followed by the analysis of the \Ha\ kinematic maps to derive rotation curves and angular frequencies (section~\ref{sec:rotcurve}). In section~\ref{sec:TW}, we outline how we apply the \tw\ method and consider the position angle uncertainties and disk coverage limitations detailed by \citet{RandWallin2004} and \citet{Zimmeretal2004}. The results are presented in section~\ref{sec:results}, compared with morphological analysis of infrared images for our sample galaxies (section~\ref{sec:morphology}), and with previous results from numerical simulations (section~\ref{sec:numsim}). Finally, the main conclusions are given in section~\ref{sec:conclusions}.

\begin{figure*}
  \begin{center}
\resizebox{\hsize}{!}{\includegraphics{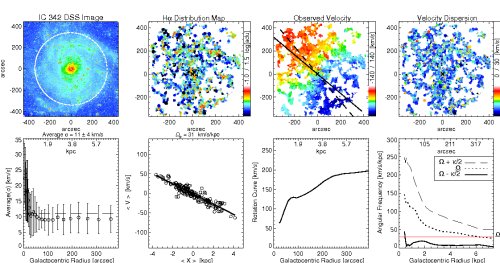}}
\resizebox{\hsize}{!}{\includegraphics{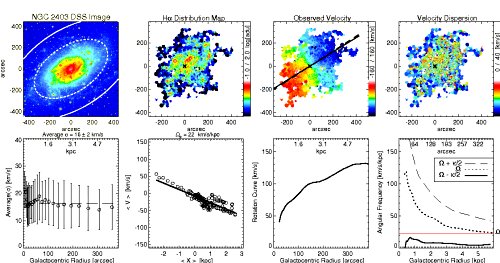}}
  \end{center}
\caption{The complete data sets for the 10 late-type spiral galaxies with $B$-band digital sky survey images included for comparison. The continuum-subtracted \Ha\ emission-line intensity, line-of-sight velocity \Vlos, and velocity dispersion $\sigma$ are found in the top row, and in the bottom row we show the galactocentric radial profiles of velocity dispersion, the \tw\ method, the rotation curve derived by means of tilted-ring and harmonic decomposition of the observed velocity field, and the angular frequency analysis. In all maps, north is up, and east is to the left. In NGC\,4519 and NGC\,6946, we mark the points from the region inside the ILR in blue colour to illustrate that they lie on a different slope than the other points. The solid ellipse outlines the 25th magnitude radius of the galactic disk, the dotted line, the $r(CR)$ of the bar or oval distortion, and the black solid ellipse oriented along the outer disk position angle, when present, outlines the position of the ILR assuming the outer disk projection parameters. The dashed line over-plotted on each velocity field outlines the kinematic major-axis \pak\ and the solid line outlines the photometric major-axis \pap\ retrieved from the RC3 catalogue. Note that the \ghafas\ data for IC\,342 only cover the central  $\approx 2$\arcmin\ radius, and therefore we only display the \fantomm\ maps here.}
\label{fig:allmaps}
\end{figure*}
\setcounter{figure}{0}\begin{figure*}
\resizebox{\hsize}{!}{\includegraphics{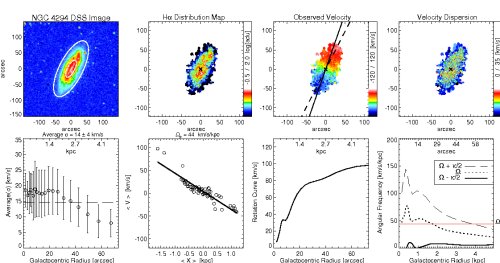}}
\resizebox{\hsize}{!}{\includegraphics{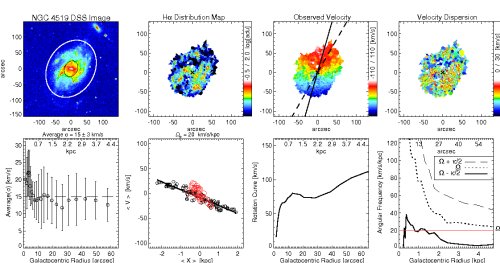}}
\caption{continued...}\end{figure*}
\setcounter{figure}{0}\begin{figure*}
\resizebox{\hsize}{!}{\includegraphics{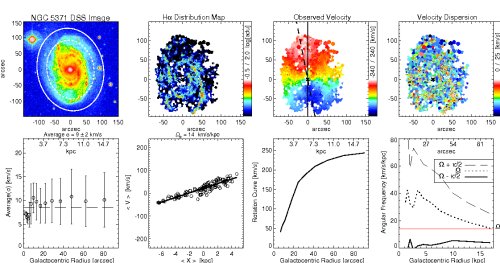}}
\resizebox{\hsize}{!}{\includegraphics{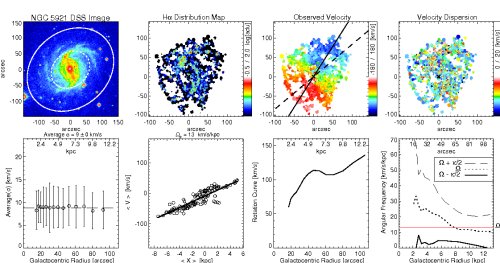}}
\caption{continued...}\end{figure*}
\setcounter{figure}{0}\begin{figure*}
\resizebox{\hsize}{!}{\includegraphics{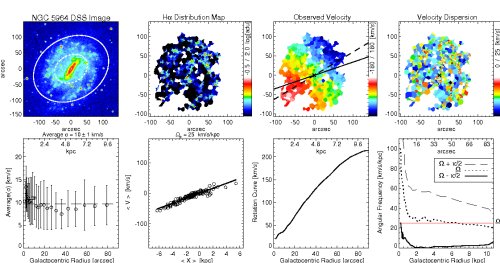}}
\resizebox{\hsize}{!}{\includegraphics{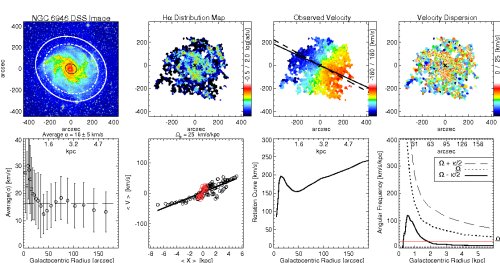}}
\caption{continued...}\end{figure*}
\setcounter{figure}{0}\begin{figure*}
\resizebox{\hsize}{!}{\includegraphics{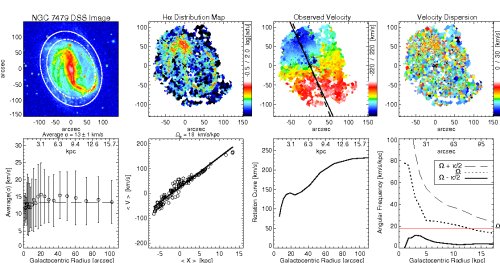}}
\resizebox{\hsize}{!}{\includegraphics{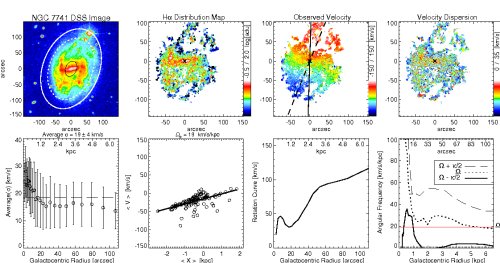}}
\caption{continued.}\end{figure*}

\begin{figure}
  \begin{center}
\resizebox{\hsize}{!}{\includegraphics{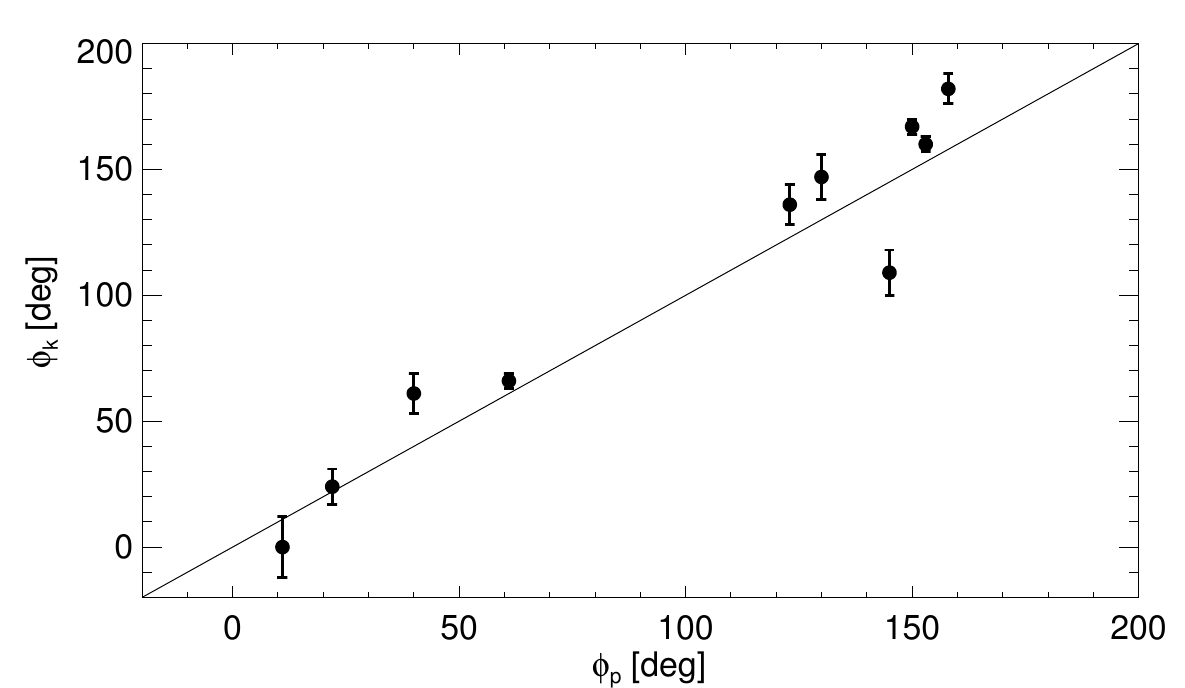}}
\resizebox{\hsize}{!}{\includegraphics{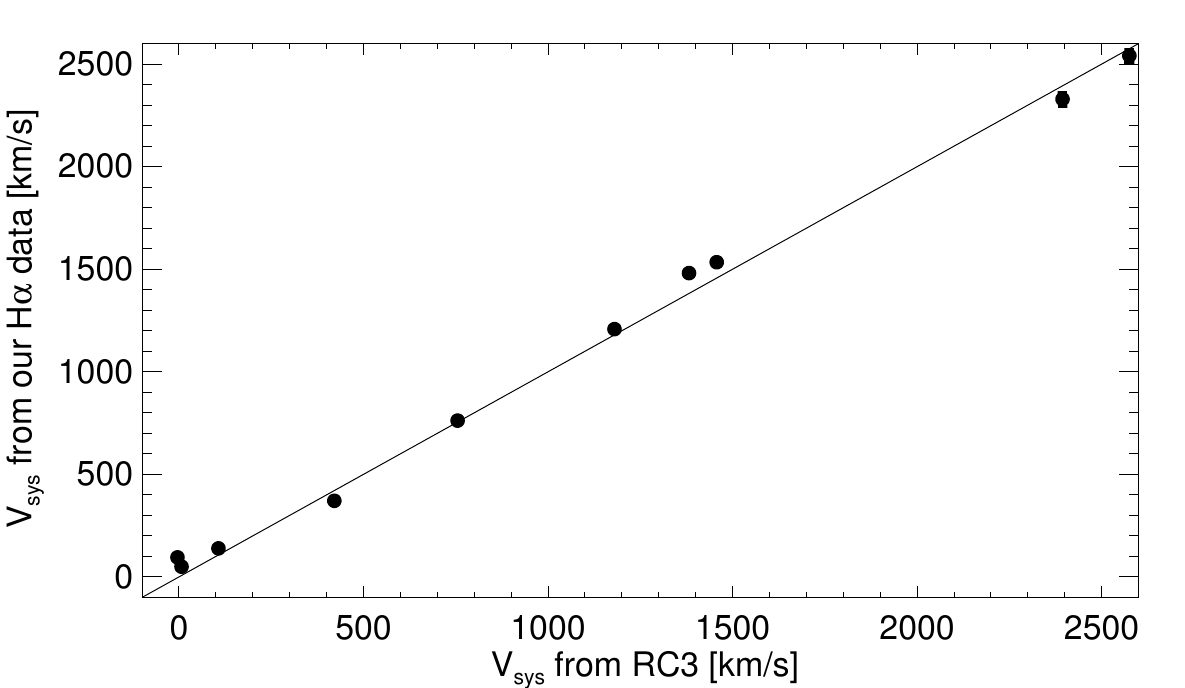}}
  \end{center}
\caption{Comparison between outer disk line-of-node position angle derived from photometry and \Ha\ kinematics (top) and \Vsys\ from the RC3 catalogue and those derived from our \Ha\ kinematics. Error bars for the kinematically derived parameters are illustrated and the diagonal line shows the 1:1 correspondence.}
\label{fig:pas}
\end{figure}

\begin{table*}
\label{tab:sampleparams}
\caption{Observed galaxies in the sample with coordinates, magnitudes, and systemic velocities from the RC3 catalogue}
\centering \tabcolsep=7pt
\begin{tabular}{lcccrrccc}
\hline \hline
 Object &R.A. (J2000)&Dec. (J2000)& Type  &$B$-band&\Vsys &Telescope& Pixel Size&Field of View\\
    \   &$h:m:s$     & $d:m:s$    & \	  & mag   &  \kms &   \   &\arcsec/pix& $\arcmin \times \arcmin$\\ \hline
IC\,342  &$03:46:49.7$&$+68:05:45$&.SXT6..&  9.10 &  -4   &OmM/WHT& 1.6/0.2 &$13.7 \times 13.7$/$3.4 \times 3.4$\\ 
NGC\,2403&$07:36:54.5$&$+65:35:58$&.SXS6..&  9.12 &  107  &OmM    & 1.6     &$13.7 \times 13.7$\\ 
NGC\,4294&$12:21:17.5$&$+11:30:40$&.SBS6..& 12.47 &  421  &CFHT   & 0.48    &$4.1 \times 4.1$\\ 
NGC\,4519&$12:33:30.6$&$+08:39:16$&.SBT7..& 12.21 &  1180 &CFHT   & 0.48    &$4.1 \times 4.1$\\ 
NGC\,5371&$13:55:40.7$&$+40:27:44$&.SXT4..& 11.34 &  2575 &CFHT   & 0.48    &$4.1 \times 4.1$\\ 
NGC\,5921&$15:21:56.5$&$+05:04:11$&.SBR4..& 11.66 &  1457 &CFHT   & 0.48    &$4.1 \times 4.1$\\ 
NGC\,5964&$15:37:36.4$&$+05:58:28$&.SBT7..& 12.60 &  1382 &CFHT   & 0.48    &$4.1 \times 4.1$\\ 
NGC\,6946&$20:34:52.1$&$+60:09:15$&.SXT6..&  9.64 &    7  &OmM    & 1.6     &$13.7 \times 13.7$\\ 
NGC\,7479&$23:04:57.2$&$+12:19:18$&.SBS5..& 11.47 &  2394 &CFHT   & 0.48    &$4.1 \times 4.1$\\ 
NGC\,7741&$23:43:54.1$&$+26:04:32$&.SBS6..& 11.66 &  755  &CFHT   & 0.48    &$4.1 \times 4.1$\\ \hline
\end{tabular}\\
{Listed are also the telescope on which \fantomm\ or \ghafas\ were mounted, the pixel size, and the field of view for each object. Although the effective field of view of \ghafas\ is $3.4 \times 3.4$ arcminutes, due to rotation of the observed field, we have used only the central $\approx 2$\arcmin\ field \citep{Hernandezetal2008}.}
\end{table*}

\section{Observations}
\label{sec:observations}
The observations were taken with the Fabry-Perot interferometer of New Technology for the Observatoire du mont M\'egantic \citep[\fantomm, ][]{Hernandezetal2003} at the 3.6m Canada-France-Hawaii Telescope (CFHT) and the 1.6m Observatoire du mont M\'egantic (OmM) (see Table~\ref{tab:sampleparams}). In addition, IC\,342 was observed with Galaxy H$\alpha$ Fabry-Perot System \citep[\ghafas, ][]{Hernandezetal2008} mounted on the 4.2m William Herschel telescope on La Palma. All details regarding the \fantomm\ observation conditions and instrument set up have been published in three papers by the Laboratoire d'astrophysique expe\'erimentale -- LAE -- at University of Montr\'eal \citep{Hernandezetal2005bhabar, Cheminetal2006, Daigleetal2006sings}. The \ghafas\ data for IC\,342 were obtained with a mean seeing of 1.7\arcsec, on January 16th 2008, during 3200 sec integrating 16 cycles of 40 channels each, with an interference order (at \Ha) $p=765\pm 1$, mean finesse $\mathcal{F}i =15.1$, and a free spectral range of 8.623 \AA.

To generate a homogeneous set of \Ha\ kinematic maps, and to gain more field coverage, we re-reduced the data using the IDL data reduction package of \citet{Daigleetal2006reduction}, with consistency checks using the procedure outlined in \citet[][ hereafter Paper~I]{Fathietal2007m74}. All data cubes were reduced and spatially re-binned to a minimum amplitude-over-noise ratio of 7. Enough frames were obtained for each galaxy to reach the background red continuum on either side of the \Ha\ line. In Fig.~\ref{fig:allmaps} we illustrate for all the galaxies, maps of the continuum-subtracted \Ha\ intensity, \Ha\ velocity, and \Ha\ velocity dispersion ($\sigma$) maps over the entire field-of-view as indicated in Table~\ref{tab:sampleparams}. For IC\,342, the \ghafas\ data only cover the central $\approx 2$\arcmin\ radius, and only the \fantomm\ maps are displayed here. Corresponding $B$-band images from the Digitized Sky Survey (DSS) are also presented for comparison with the Fabry-Perot maps.

\section{Deriving Rotation Curves and Velocity Dispersion Profiles}
\label{sec:rotcurve}
We quantify the observed velocity fields by using the tilted-ring method combined with the harmonic decomposition formalism \citep[e.g., ][]{Schoenmakersetal1997, Fathietal2005}. We assume that circular rotation is the dominant feature and that our measurements refer to positions on a single inclined disk, and divide the velocity field in concentric rings and fit the systemic velocity \Vsys\ and disk geometry (position of the centre ($x_0, y_0$), inclination $i$, line of nodes position angle $\phi$). We fit the circular and non-circular motions within each ring simultaneously. The inclination is fixed at the value taken from the RC3 catalogue \citep{deVaucouleursetal1991} and the projected position of the dynamical centre, \Vsys, and kinematic line of nodes (\pak\ ) are fitted iteratively \citep[see also ][ ]{Begeman1987}. After each iteration, one of the parameters is fixed to the average value of all rings. 

We first fix the dynamical centre to the average $x_0$ and $y_0$ value for all rings, second, we fit $V_\mathrm{los} = V_\mathrm{sys} + \sin(i)\,[c_1 \cos(\phi_k) + s_1 \sin(\phi_k)]$ for each ring. After calculating \pak\ the average value of the line of nodes for all rings, we derive the rotational and radial velocity components together with the higher harmonic terms simultaneously (up to and including the third-order term). In the case when higher terms are absent, the harmonic parameters are $c_1$ and $s_1$ which include \Vrot\ and \Vrad. This procedure minimises the contribution from symmetric streaming motions caused by bars or spiral arms on the derived rotation curves \citep[see ][ and Paper~I, for a more details]{Schoenmakersetal1997,Fathietal2005}.

To investigate how the choice of data parameters in the data reduction procedure affects the final results, we applied this procedure to velocity fields with minimum amplitude-over-noise values varying between 1 and 15. We find that the differences in the results from these data cubes could be used to derive reasonable error estimates, since the kinematic parameters derived from these different maps vary within a range larger than the statistical errors derived for each map (see Table~\ref{tab:kinparams}). The disk geometry that best fits each observed velocity field is then used to derive \Vrot\ and to section the $\sigma$ maps into the corresponding tilted rings. For each ring, we average all values from individual pixels to construct the $\sigma$ profiles presented in Fig.~\ref{fig:allmaps}. 

Our kinematically derived \Vsys\ values are similar to the systemic velocities listed in the RC3 catalogue (see Fig.~\ref{fig:pas}), however, since almost all galaxies are subject to significant amounts of non-cosmological velocities, we cannot use the \Vsys\ to directly derive cosmological distances. For all galaxies, we have been able to find, in the literature, accurate distance measurements from radio source studies, Cepheid variables, the Tully-Fisher relation, or corrected for the Virgo cluster perturbation \citep[][]{Tully1988, FreedmanMadore1988, Gavazzietal1999, Daleetal2000, Kennicuttetal2003}. We use these distances throughout our analysis.

\begin{table}
\caption{Uncertainties for the dynamical centre position ($x_0(err), y_0(err)$, east and north-ward respectively), inclination, \pak\ values given in north-east direction, and distances and arcsec-to-parsec scales as described in section~\ref{sec:rotcurve}.}
\centering \tabcolsep=4pt
\begin{tabular}{lcccrrr}
\hline \hline
 Object  &$x_0(err)$&$y_0(err)$& incl. & \pak     &Distance & Scale \\  
    \    &\arcsec   &\arcsec   & deg   & $ deg $	&  Mpc    & pc/\arcsec \\ \hline
IC\,342   &1.3	   &3.3      & 25    &$ 61\pm 8$&   3.9   & 19  \\   
NGC\,2403 &2.8	   &1.9      & 60    &$136\pm 8$&   3.2   & 16  \\
NGC\,4294 &1.5	   &1.1      & 65    &$160\pm 3$&  14.0   & 68  \\
NGC\,4519 &0.5	   &1.6      & 36    &$167\pm 3$&  15.1   & 73  \\
NGC\,5371 &1.8	   &1.1      & 48    &$  0\pm12$&  37.8   & 183 \\
NGC\,5921 &1.6	   &5.7      & 45    &$147\pm 9$&  25.2   & 122 \\
NGC\,5964 &1.2	   &2.4      & 20    &$109\pm 9$&  24.7   & 120 \\
NGC\,6946 &1.0	   &1.3      & 24    &$246\pm 3$&   6.5   & 32  \\
NGC\,7479 &1.1	   &1.2      & 42    &$ 24\pm 7$&  32.4   & 157 \\
NGC\,7741 &2.6	   &0.3      & 48    &$  2\pm 6$&  12.3   & 60  \\ \hline
\end{tabular}
\label{tab:kinparams}
\end{table}

\begin{figure*}
\resizebox{\hsize}{!}{\includegraphics{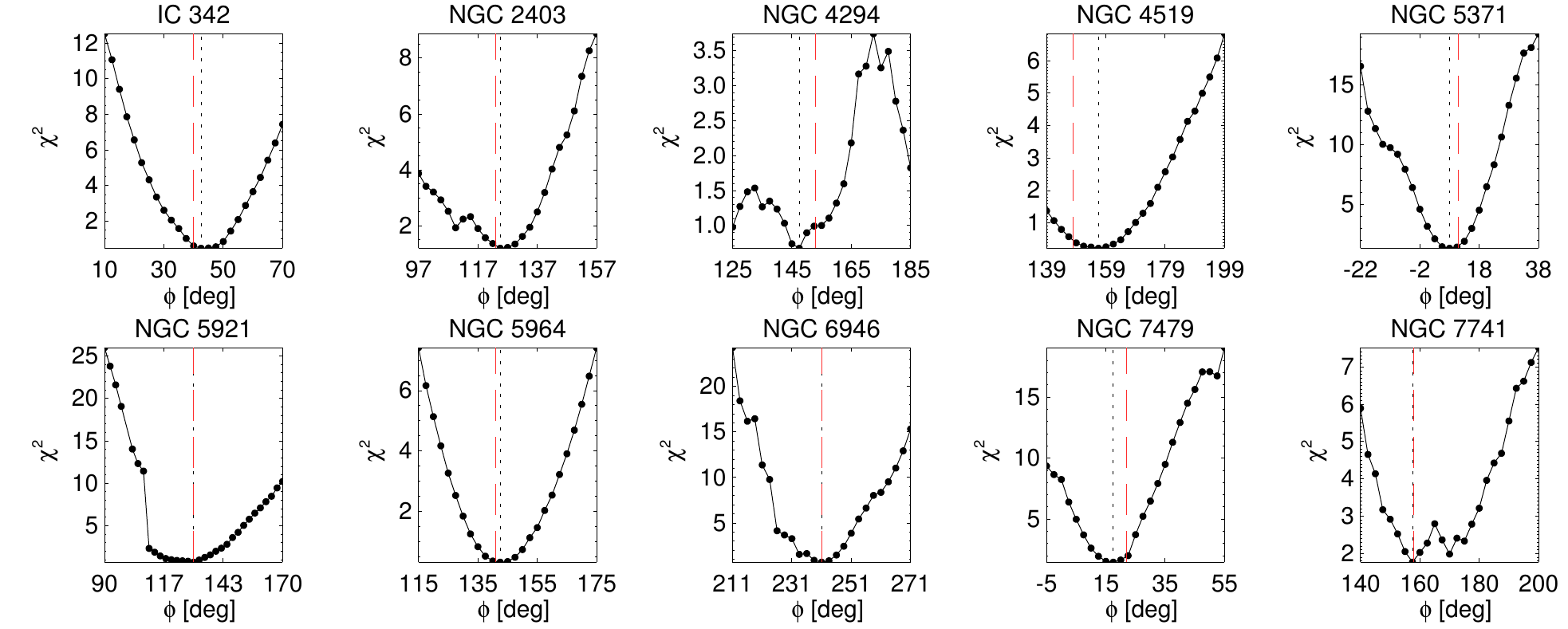}}
\caption{Reduced $\chi^2$ values for the linear fits to the \tw\ diagrams along different line of nodes. The vertical dotted line indicated the angle $\phi$ where the minimum $\chi^2$ value is obtained, and the dashed (red) line shows the best estimate for disk \pap\ as described in section~\ref{sec:pa}. For all \Vlos\ measurements, a constant velocity error of 10 \kms\ has been used to calculate the reduced $\chi2$ values.}
\label{fig:PA_chi2}
\end{figure*}

\begin{figure*}
\resizebox{\hsize}{!}{\includegraphics{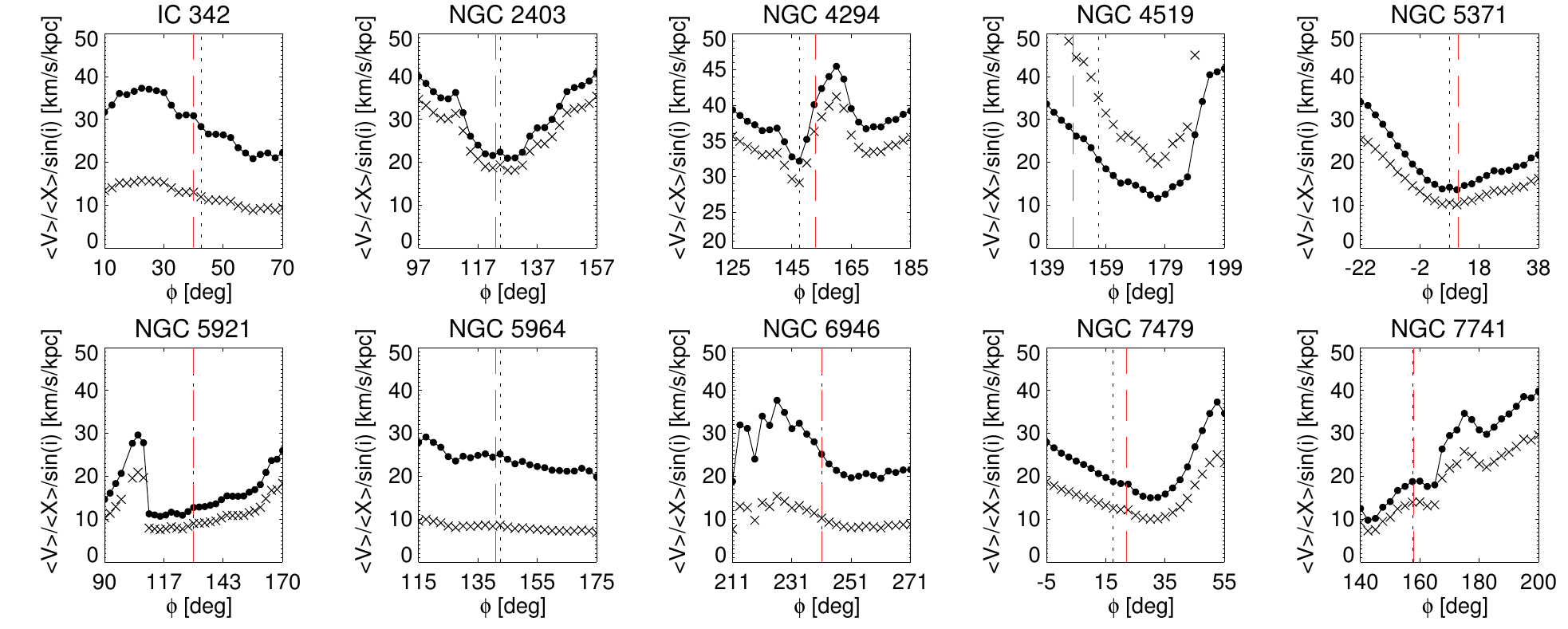}}
\caption{Same as Fig.~\ref{fig:PA_chi2}, but here, the $y$-axis shows the derived $\langle V\rangle/\langle X\rangle/ \sin(i)$ (filled symbols) along each angle. The ``crosses'' show the same quantity multiplied by $\sin(i)$ to compare the values independently of the assumed inclination.}
\label{fig:PA_Op}
\end{figure*}

\subsection{Position Angle of the Outer Disk}
\label{sec:pa}
When applying the \tw\ method, the slits must be placed along the line of nodes, which can be approximated by the kinematic and/or photometric major-axes. 

The orientation of the kinematic major-axis  derived from the quantification scheme described in section~\ref{sec:rotcurve} is the result of the second iteration in the procedure for deriving \Vrot, i.e., after determining and fixing the position of the dynamical centre ($x_0, y_0$) and \Vsys. Similar to the standard tilted-ring method, the algorithm minimises the expression $\chi^2 = V_\mathrm{los}^2 - (c_1^2 + s_1^2)$, which can be seen ideally in cases when \Vlos\ is the result of a circular cosinusoidal term ($c_1$) and a radial sinusoidal term ($s_1$). In the presence of weak non-circular motions caused by weak perturbations or large beam smearing effects as in radio observations, this approximation is reasonable. However, for our data, where local velocities from strongly star forming regions as well as significant flows along the bars in the objects could significantly skew the isovelocity contours towards the bar axis \citep[e.g., ][]{Duval1977, Huntley1978, Buta1987}, we are bound to derive misplacements between the kinematic line of nodes position angle \pak\ and the photometric counterpart \pap. One step forward in treating this problem is the expansion of the minimisation routine to the expression  $\chi^2 =  V_\mathrm{los}^2 - \Sigma_{m=1}^3 (c_m^2 + s_m^2)$. However, patchiness of observed field, the presence of strong dust content internal to the observed galaxy, and asymmetric drift cause considerable computational complications which are beyond the scope of this paper \citep[e.g., ][]{Weijmansetal2008}. 

We derive the outer disk line of nodes position angle (\pap\ ) from the near infrared images presented in section~\ref{sec:ellipticities} by using the average position angle for the outer isophotes over a 2--4 kpc range in radius. We compare the values with those obtained from the RC3 catalogue as well as previous studies of our sample galaxies, and find that in most cases, the photometrically derived values are in fair agreement (see Table~\ref{tab:PAvalues}), however, for IC\,342, NGC\,5921, and NGC\,6946, the \pap\ values from the literature are based on comprehensive analysis, we use these angles for deriving the pattern speeds.

Furthermore, we compare the $\chi^2$ values for the linear fits to the $\langle V\rangle$, $\langle X\rangle$ pairs, and as it is expected that the \tw\ method delivers the least scatter along the correct line of nodes, we use the $\chi^2$ values to assess the \pap\ we have decided to use. Fig.~\ref{fig:PA_chi2} illustrates the reduced $\chi^2$ values for a range of position angles, where the minimum $\chi^2$ values correspond to position angles almost identical to or to within 10 degrees from \pap. This figure also shows that extracting  ($\langle V\rangle$, $\langle X\rangle$) pairs along offset position angles of $\pm 10$ degrees still delivers reasonably linear fits to the $\langle V\rangle$, $\langle X\rangle$ diagrams such as those shown in Fig.~\ref{fig:allmaps}.

We use these uncertainties to measure differences in the \Op\ by changing the orientation of the nominal slits. The $\langle V\rangle / \langle X\rangle /\sin(i)$ values illustrated in Fig.~\ref{fig:PA_Op} show that for \pap\ uncertainties of $\pm 10$ degrees, we can bracket the \Op\ values to within the limits shown listed in Table~\ref{tab:patternspeeds}. Allowing the slit widths to vary between 1 and 10 pixels implies additional differences in the derived \Op\ of the order of 10\%.

\begin{table*}
\caption{Disk line of node position angles from different sources given in degrees from north to east, and the \pap\ value which we use throughout the paper.}
\centering \tabcolsep=7pt
\begin{tabular}{lrrrrcl}
\hline \hline
Galaxy & $\phi$ (RC3) & \pak\ & $\phi$ (NIR) & $\phi (\chi^2_{min})$ & \pap\ & $\phi $(literature) \\[1mm]  \hline
IC\,342    &$\dots$ &   61   &	  48   &    42.5   & \;\; {\bf 40}\;\;	&  39, 40, 87\;\citep{Newton1980, Crosthwaite2002, ButaMcCall1999} \\[1mm] 
NGC\,2403  &   127  &   136  &	  123  &   124.5   & \;\;{\bf 123}\;\;  &  124\;      \citep{Barberaetal2004} \\[1mm]  
NGC\,4294  &   155  &   160  &	  153  &   147.5   & \;\;{\bf 153}\;\;  &  151, 148\; \citep{Patureletal2000, Distefanoetal1990} \\[1mm] 
NGC\,4519  &   169  &   167  &	  148  &   158.5   & \;\;{\bf 148}\;\;  &  152, 145\; \citep{Patureletal2000, Rubinetal1999} \\[1mm] 
NGC\,5371  &     8  &   0    &	  11   &     8.0   & \;\; {\bf 11}\;\;	&  -4, 17\;   \citep{Patureletal2000, Barberaetal2004} \\[1mm] 
NGC\,5921  &   130  &   147  &	  97   &   130.0   & \;\;{\bf 130}\;\;  &  150, 130\; \citep{Patureletal2000, Knapenetal2006} \\[1mm]
NGC\,5964  &   145  &   109  &	  141  &   142.5   & \;\;{\bf 141}\;\;  &  149\;	    \citep{Springobetal2007} \\[1mm] 
NGC\,6946  &$\dots$ &   221  &	  231  &   241.0   & \;\;{\bf 241}\;\;  &  240\;      \citep{Springobetal2007} \\[1mm] 
NGC\,7479  &    25  &   24   &	  22   &    17.5   & \;\; {\bf 22}\;\;	&  29, 22\;   \citep{Patureletal2000, Laineetal1998}\\[1mm]
NGC\,7741  &   170  &   182  &	  158  &   157.5   & \;\;{\bf 158}\;\;  &  162\;      \citep{Patureletal2000}\\[1mm] \hline
\end{tabular}
\label{tab:PAvalues}
\end{table*}

\begin{figure}
\includegraphics[width=0.5\textwidth]{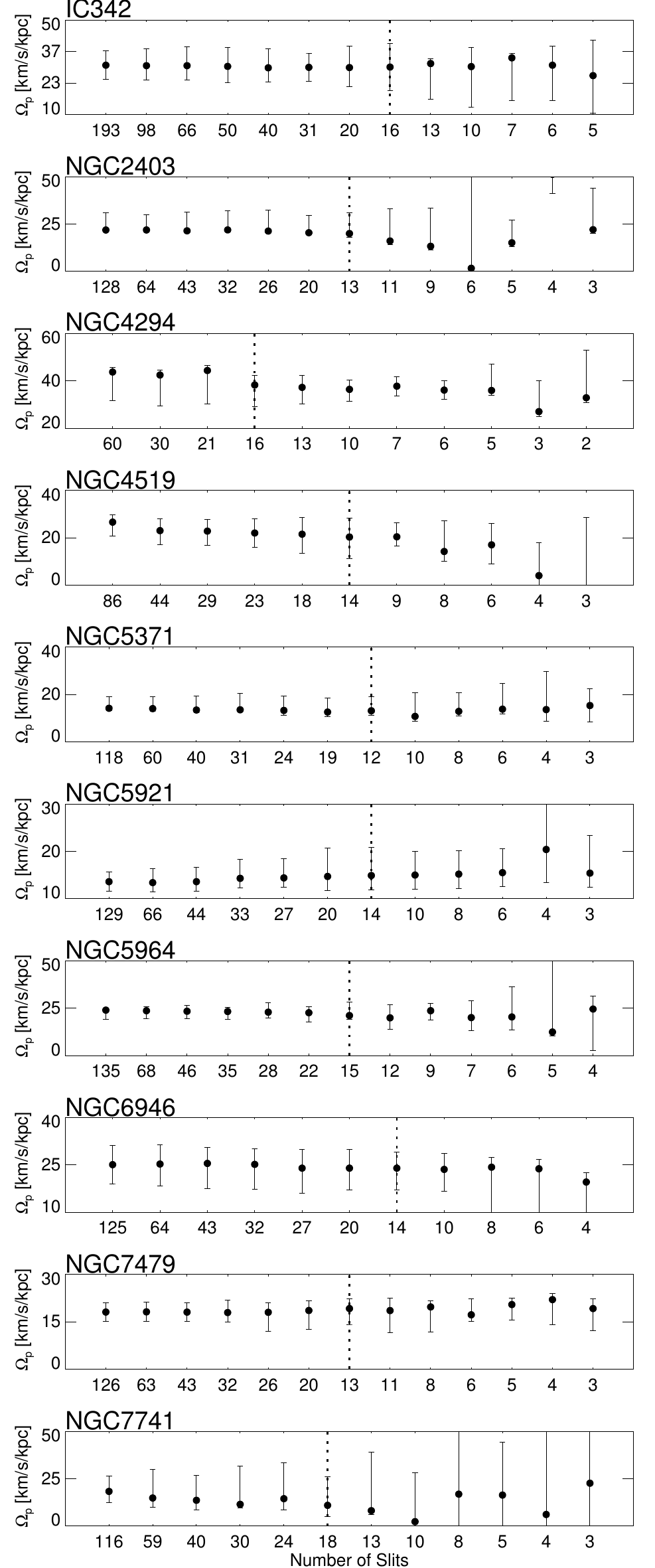}
\caption{Comparison between \Op\ and the varying width of the pseudo slits. For a comparison between our work and the \CO\ work, the vertical dotted line indicates when a total of $\approx 15$ slits covers the entire observed field (which is the typical number of slits used in the previous studies). Lowering the resolution (or using wider slits) delivers \Op\ uncertainties in agreement with the findings of \citet{Debattista2003} and \citet{RandWallin2004}.}
\label{fig:slitwidth}
\end{figure}

Previous analysis of the \Op\ uncertainties due to incorrect slit orientations have been carried out by \citet{Debattista2003} and \citet{RandWallin2004}, who found that $\approx 5$ degrees position angle (PA) offset is sufficient to cause \Op\ values that differ by up to 30\%. In Fig.~\ref{fig:slitwidth}, illustrated as error bars, we show the difference in the measured \Op\ between an estimate with the correct line of nodes, and an estimate using an incorrect line of nodes, displaced by 10 degrees, as a function of the number of slits used in these estimates. The figure shows that this difference grows quickly as the number of slits falls. When the number of slits has fallen to around 15, matching this condition, the differences have grown to 30\%, and thus consistent with the findings of \citet{RandWallin2004}. However, we note that in our derivations of the \Op\ we use over 100 slits per galaxy.

\begin{figure}
\includegraphics[width=0.5\textwidth]{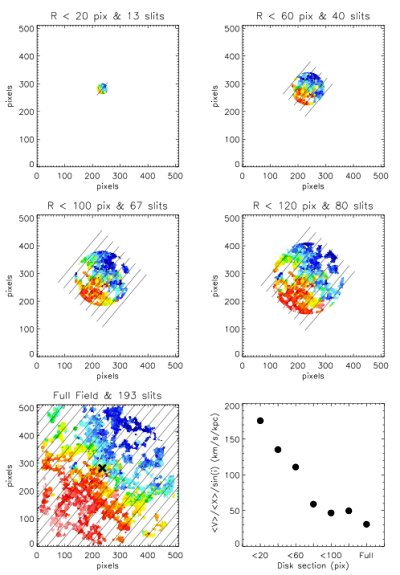}
\caption{Illustrating the case of IC\,342 as an example: Sections of the disk used to explore the radial dependency of the \tw\ method to investigate whether the depth of the observations could deliver erroneous values. The diagonal lines indicate the orientation of the fiducial pseudo-slits (one in every ten slits is marked), and the cross in the bottom left panel shows the position of the dynamical centre. Written on each panel, are the radial extent and number of extracted slits.}
\label{fig:radialdependency}
\end{figure}

\section{Applying the \tw\ Method}
\label{sec:TW}
For each galaxy our two dimensional velocity field allows us to select a set of pseudo-slits which we can simply call "slits". We select these slits such that they are three pixels wide and parallel to the line of nodes of the outer disk. For each independent point along a slit we determine the  luminosity-weighted value of  \Vlos\ - \Vsys\ and the luminosity-weighted position $\langle X \rangle$, using as well the continuum-subtracted \Ha\ or the stellar continuum surface brightness maps from the data cube as our weighting function. Checking that there are measurable velocity data from more than 50 pixels along any given slit, we then compute the Tremaine-Weinberg integral along that slit, giving us a single point in our plot (see Fig.~\ref{fig:allmaps}). The resulting plots have far more points in them than conventional plots obtained using the long-slit spectrum of the stellar component or \CO\ observations, because we are able to use far more slits; a typical number for the galaxies analysed in this article is $>100$. As we can see from Fig.~\ref{fig:allmaps}, ``clean'' linear fits are found for almost all the objects. 

Even though all the \Ha-emitting gas does not satisfy assumption 3, we note two points in support of using the \tw\ method with our data:

Firstly, from the maps displayed in continuum-subtracted \Ha\ in Fig.~\ref{fig:allmaps} (as well as those discussed in Paper~I), a major fraction of the pixels ($> 80\%$) in our maps are due to the diffuse ionised gas component, which is much less contaminated by the local dynamical effects such as effects of supernovae and winds from individual \HII\ regions. Deep \Ha\ observations of NGC\,7793 \citep{Dicaireetal2008}, and imaging of the SINGS sample \citep{Oeyetal2007} further confirm the contribution of the diffuse component. Combined with the typical star formation rates, $<1$ \Myr\ for half of our sample,  estimated by \citep{Kennicuttetal2003}, \citet{MartinFriedli1997}, and \citet{GonzalezDelgado1997}, we expect that only a minor fraction of the velocities in the observed fields are affected by local effects such as winds from star forming complexes.

Secondly, the gaseous motions which do not share the velocity field of the global pattern may, in large part, cancel due to global symmetries (arms, bar) or local symmetry (the expansion of individual \HII\ regions). The bi-symmetric streaming motions along the bar and the arms must cancel quite well in order to restrict the mean gas inflow rates towards the centre of a galaxy. Local interruptions to continuous flow, while real, are on average circularly symmetric in the plane of the galaxy and any departure from regularity which might cross out continuity will cancel because there is sufficient symmetry in these departures. Example for such departures include star formation at the ends of a bar or two regular arms littered with \HII\ regions. Thus, the resulting values for the pattern speed are not necessarily less accurate than those obtained using the stellar component, especially as the latter are subject to relatively poor signal-to-noise conditions, and would need exposures more than one order of magnitude longer to give similar signal-to-noise ratios to those we find here.

Regarding any possible effects of the spiral density wave in triggering star formation, there are two points to note. (I) Using the stellar continuum as our weighting function (see Table~\ref{tab:patternspeeds}, and further discussion on this point in section~\ref{sec:results}) systematic effects from \Ha\ due to the presence of newly formed stars are ruled out. Furthermore, when we use the \Ha\ emission intensity for weighting the velocities, columns 3 and 4 in Table~\ref{tab:patternspeeds} show that the resulting \Op\ do not differ too greatly from those obtained with the red continuum measured from the Fabry-Perot data cubes. The typical difference is 3 \kmskpc\ which fits within the typical uncertainties for our measurements. This suggests that star formation triggering effects, although locally important, do not change the overall weights applied to the velocities. (II) We consider velocity deviations due to density waves by making sure that we derive consistent values when we move along a given slit or across the disk. This strongly suggests that such velocity deviations do cancel to first order.

\begin{table*}
\caption{Pattern speeds (in \kmskpc) derived using continuum-subtracted \Ha\ maps and continuum maps. The values in the brackets list the differences in the derived pattern speeds when varying the pseudo slit orientations by $\pm 10$ degrees, and they are of the same order for the columns 2 and 5.}
\centering \tabcolsep=10pt
\begin{tabular}{lccccl}
\hline \hline
 Object   &\Op\ along \pak & \Op\ along \pap\ (\Ha) & $\Omega_p\times\sin(i)$ along \pap\ & \Op\ along \pap\ (continuum) & \Op\ Literature\\  
(1) & (2) & (3) & (4) & (5) & (6)\\[1mm] \hline

IC\,342       &36 & 31    (-5, 5)&  13 &  32    & 10-40    \\[1mm]  
NGC\,2403     &22 & 22    (-1, 9)&  19 &  25    & 10.5     \\[1mm] 
NGC\,4294     &39 & 43   (-12, 2)&  39 &  37    & \dots    \\[1mm] 
NGC\,4519     &13 & 20    (-8, 6)&  12 &  33    & \dots    \\[1mm] 
NGC\,5371     &16 & 14    (-1, 5)&  10 &  17    & \dots    \\[1mm] 
NGC\,5921     &13 & 13    (-2, 2)&   9 &  10    & 15-20    \\[1mm] 
NGC\,5964     &23 & 25    (-5, 1)&   9 &  24    & \dots    \\[1mm] 
NGC\,6946     &21 & 25    (-6, 6)&  10 &  27    & 22-66    \\[1mm] 
NGC\,7479     &18 & 18    (-3, 3)&  12 &  15    & 27-100   \\[1mm] 
NGC\,7741     &33 & 19    (-6, 8)&  14 &  18    & 30-145   \\[1mm] \hline
\end{tabular}\\
{Columns 2, 3, and 4: \Op\ derived along \pak\ and \pap\ using continuum-subtracted \Ha\ emission-line intensities, and inclination-corrected value, and column 5: using the continuum maps (described in section \ref{sec:observations}) instead of the \Ha\ maps. References for the literature values (column 6) are further discussed in section~\ref{sec:notes}.}
\label{tab:patternspeeds}
\end{table*}

\begin{figure*}
\resizebox{\hsize}{!}{\includegraphics{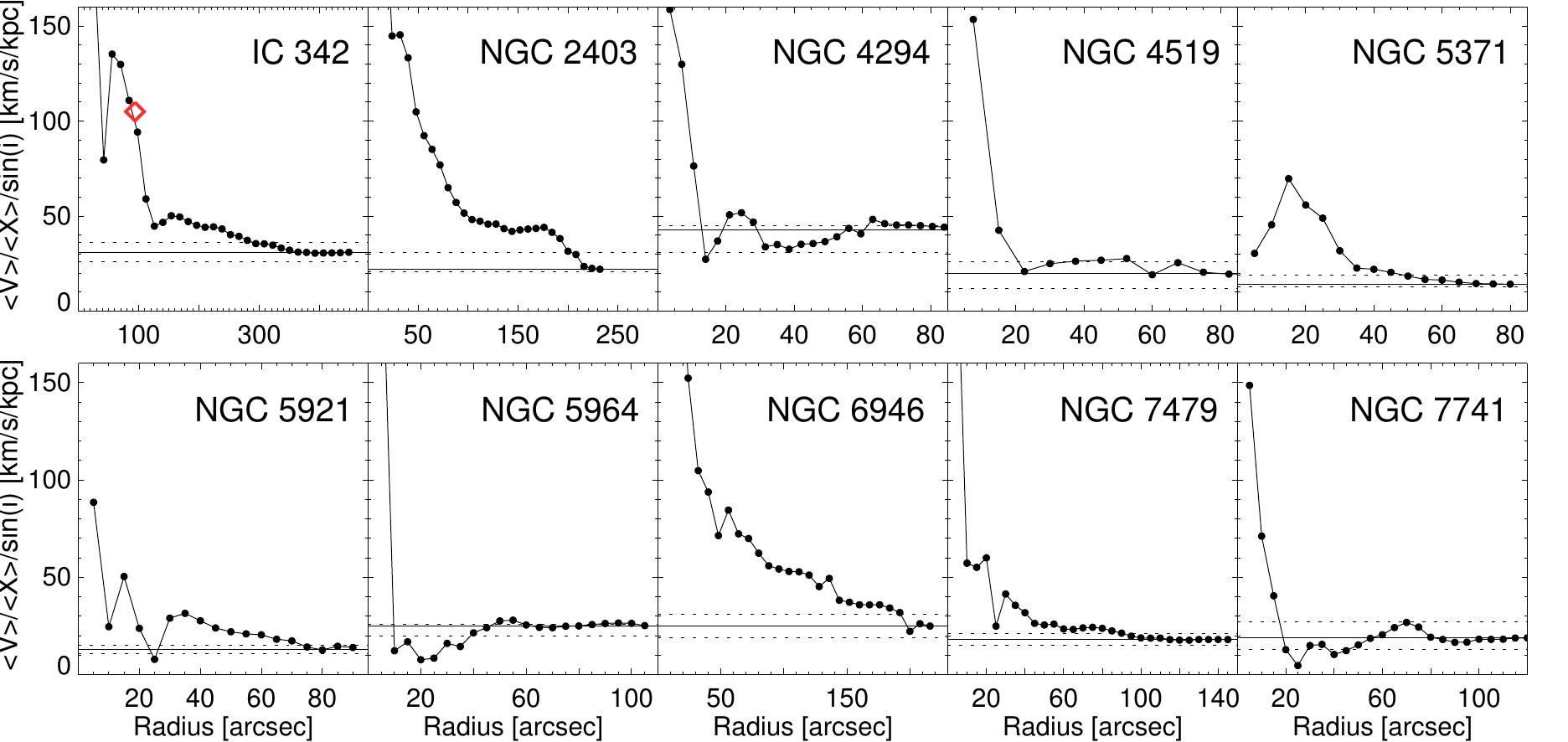}}
\caption{The variation of the $\langle V\rangle / \langle X\rangle /\sin(i)$ values, along the \pap\ angle, using different sections of the disk as described in section~\ref{sec:radius}. The diamond in the IC\,342 panel shows the value derived using the inner 2\arcmin\ field of \ghafas, and the horizontal solid line indicates the pattern speed derived using the entire observed field with upper and lower limits indicated by the dashed lines (column 3 of Table~\ref{tab:patternspeeds}).}
\label{fig:radialOpvalues}
\end{figure*}

\subsection{Changing the limits and the weights when calculating the \tw\ integrals}
\label{sec:radius}
The ideal form of the Tremaine-Weinberg method uses the luminosity-weighted velocities and the distances in the full radial range of the galaxy disk, i.e. the integrals range from $-\infty$ to $\infty$. Observations, on the other hand, cover only a limited radial range, and ours reach typically the $r_{25}$ radius of the galaxy disk. It is thus important to see whether our observations are deep enough so that they do not yield incorrect \Op.

We note that the \tw\ method gives pattern speeds strictly when the limits of the integrals reach well beyond the bar.  However, in this test, we restrict the radial range of integration in progressive steps, starting from an innermost radius of 10 pixels and increasing in steps of 10 pixels until the entire observed disk is covered (see Fig.~\ref{fig:radialdependency}). Each derivation yields a different $\langle V \rangle$/$\langle X \rangle$ value, treated as if it were a value for \Op\ (see Fig.~\ref{fig:radialOpvalues}). As shown in the detailed analysis by \citet{Zimmeretal2004} these values decrease asymptotically as the disk coverage increases. In particular the differences between the outermost values of these trials and the values obtained using the complete disks are much smaller than those between the values obtained limiting the field to a few tens of pixels and those found when using the full disk region. Furthermore, we confirm that by using all the pixels inside the slit (i.e., without the radial truncation as shown in Fig.~\ref{fig:radialdependency}), the derived \Op s are much more stable. This is in support of the result by \citet{Zimmeretal2004} who found that most of the variation in the derived pattern speed occurs within an inner radius which they set to 100\arcsec\ for their observations.

A pattern speed derived over a radial range significantly shorter than the bar length will not give a meaningful value since the integrals do not converge, so we do not wish to justify here any variations (or indeed absence of variations) calculated within these ranges. Moreover, as found by \citet{Shettyetal2007} for M\,51, quasi-steady state continuity is not satisfied by a single valued pattern speed, so it is important to analyse the derived \Op\ variations to make sure they converge on a unique value, which can then be taken as the main value for a single and dominant $m=2$ perturbation \citep{Henryetal2003}. In our work we have used integrations over the full observed field for all galaxies to derive the pattern speeds presented in Table~\ref{tab:patternspeeds} and Fig.~\ref{fig:allmaps}.
 
In an ongoing study of the \tw\ method based on N-body + SPH numerical simulations carried out by Isabel P\'erez \citep[P\'erez et al. in preparation, see also ][]{Beckmanetal2008} we find that in the case of one constant pattern speed in a disk, scans of untruncated pseudo-slits in the inner regions as well as outer regions, all recover the input \Op\ of the simulations. When truncating the slits to cut the outer sections of the disk (compare with Fig.~\ref{fig:radialdependency}), the \tw\ integrals applied to inner truncated slits result in different $\langle V \rangle$/$\langle X \rangle$ values when compared with using the full simulated galaxy. This is a clear demonstration of the necessity of using full slit coverage and deep observations (even in the inner regions) due to the non-converging integrals. However, as expected, we have found that as soon as the radial disk section is comparable with or larger than the size of the bar, the \Op\ input into the simulations can be recovered.  This feature is shared by real data as well as by numerical simulations.

Furthermore, we have applied the \tw\ method using the simulated velocity field weighted by the simulated density maps as well as the simulated star-formation maps and found that although using the density maps provide more accurate results, the star-formation maps also deliver comparable (to within the errors) pattern speeds,  implying that the \Ha\ velocities are reliable when applying the \tw\ method. For the galaxies presented here, Table~\ref{tab:patternspeeds} shows that the \Ha\ and continuum deliver comparable \Op\  (note that the continuum maps used for this exercise also come from our data cubes). The exception is NGC~4519, where the continuum image assigns higher weights to the velocities inside the bar, and thereby introduces significant difference between the \Op\ derived using the continuum and \Ha\ maps. The higher pattern speed of the bar, using the continuum image delivers a higher total \Op.

In the presence of multiple structures with different pattern speeds, when deriving the $\langle V \rangle$/$\langle X \rangle$ for an inner component, such as an inner bar inside a large bar, by applying the notional slits on the inner parts of a two-dimensional map, the majority of the pixels will belong to the disk region outside the bar. These pixels will therefore affect the derived pattern speed of the inner structure. This effect was tested by e.g., \citet[][ their Fig.~7 ]{Zimmeretal2004} who found that once the slit coverage is comparable with the size of the main bar, the \tw\ method delivers reliable \Op s.  Moreover, cutting the disk in radial bins seems to reveal the presence of multiple pattern speeds when the disk section is comparable to structure with a correspondingly distinct pattern speed. In Fig.~\ref{fig:allmaps}, we illustrate the $\langle V \rangle$, $\langle X \rangle$ points from the inner regions of two galaxies where this effect can be seen. However, as the derivation of secondary pattern speeds is more complicated \citep[see e.g., ][]{Meidtetal2008}, this test is to be used with caution in those cases.

Finally, we confirm that the \Op\ differences using the continuum images instead of the \Ha\ surface brightness maps, does not significantly change the location of the resonances throughout our analysis.

\section{Comparison With Morphology}
\label{sec:morphology}
Morphologically, the bar radius r(bar) is commonly determined by analysing ellipse fitting profiles \citep[e.g., ][]{Athanassoula1992, Wozniaketal1995}, or the two-fold symmetric component in a Fourier decomposition of the isophotes \citep{ElmegreenElmegreen1985ApJ, Buta1986I}. We have used both methods to verify the validity of the Corotation Radius $r(CR)$ values that we have derived from the \tw\ method applied to the \Ha\ velocity fields.

\subsection{Ellipticity Profiles}
\label{sec:ellipticities}
Empirical studies by \citet{Erwin2005} and simulations by \citet{Martinez-Valpuestaetal2006} and \citet{Gadottietal2007} have shown that in systems with single bars, careful examination of ellipticity ($\epsilon = 1 - b/a$) profiles can deliver a reliable $m=2$ bar radius.  For our sample galaxies, we analyse near infrared $K_s$-band images from the 2MASS catalogue \citep{Jarrettetal2000} and deep 3.6 $\mu$m images from the Spitzer archives (PI:Fazio, PI:Kennicutt, PI: Kenney) in order to derive the bar length. For the galaxies where Spitzer data were not available, we retrieved deep $H$-band image from the 2.2m Calar Alto telescope (NGC\,4519; PI: Boselli) and deep $K$-band images from the 4.2m William Herschel Telescope \citep[NGC\,5921 and NGC\,5964; ][]{Knapenetal2003}.

We detect the presence of a bar within a galaxy disk from the ellipticity and line of nodes position angle $\phi$ profiles traced by the galaxy isophotes \citep{Jedrzejewski1987}. The transition from the (round) bulge-dominated centre to the disk is typically characterised by an increase in ellipticity. Following \citet{Martinez-Valpuestaetal2006}, we define the semi-major axis length of the bar ($r(bar)_\epsilon$) as the radius where the ellipticity value has fallen by $15\%$ \citep[empirically derived from the numerical simulations by ][]{Martinez-Valpuestaetal2006} from the radially outermost local maximum accompanied by no significant change in $\phi$ (see Fig.~\ref{fig:morphology}). It is important that no position angle variation is detected around this radius, as twists could imply erroneous ellipticity diagnostics.

We compare the $r(CR)$ derived from kinematic analysis with the bar radius derived from the images (see Table~\ref{tab:CRoverLength}), and find that the ratio $r(CR)/r(bar)_\epsilon$ agrees with the predictions from numerical simulations \citep{Athanassoulaetal1990, ONeillDubinski2003}. This result further suggests that peak ellipticities could underestimate the radial extent of the bar, though not by a large factor, since the ellipticity profiles shown in Fig.~\ref{fig:morphology} illustrate that the radius where ellipticity value has fallen by $15\%$ is usually not located far from the radius of peak ellipticity. Furthermore, for almost all galaxies, we derived comparable ellipticity profiles using the 2MASS $K_s$-band images. However, these images are not deep enough to deliver good quality ellipticity profiles at large radii.

\begin{table*}
\caption{Corotation radius  $r(CR)$ and bar lengths derived form ellipse fitting ($r(bar)_\epsilon$) and Fourier analysis of the near infrared images with uncertainties calculated by allowing for the \Op\ to vary within the limits listed in Table~\ref{tab:patternspeeds}. In NGC\,4519, $r(CR)$ falls outside our observed field, and for NGC\,5371, we have measured the $r(CR)$ for the spiral arms, hence no ratios calculated.}
\centering \tabcolsep=16pt
\begin{tabular}{lllllll}
\hline \hline
 Object     &$r(CR)$	  &  $r(CR)$ & $r(bar)_\epsilon$ & $r(CR)/r(bar)_\epsilon$ & $r(bar)_F$ & $r(CR)/ r(bar)_F$ \\	 
 \          &\arcsec	  &  kpc  	       & kpc   & \    & kpc	& \ \\ \hline \\	
IC\,342     &$344_{-79}^{+26}$  & $6.5_{-1.5}^{+0.5}$    &5.8  &$1.1_{-0.3}^{+0.1}$    &7.2   &$0.9_{-0.2}^{+0.1}$  \\[3mm]
NGC\,2403   &$390_{-32}^{+?}$   & $6.2_{-0.5}^{+?}$      &7.8  &$0.8_{-0.1}^{+?}$      &6.6   &$0.9_{-0.1}^{+?}$    \\[3mm]
NGC\,4294   &$29_{-6}^{+4}$	  & $1.9_{-0.4}^{+0.3}$    &1.7  &$1.1_{-0.2}^{+0.2}$    &2.5   &$0.8_{-0.2}^{+0.1}$  \\[3mm]
NGC\,4519   &$\dots$  	  & $\dots$	       &4.7  &$\dots$	       &4.7   &$\dots$	      \\[3mm]
NGC\,5371   &$\dots$            & $\dots$	       &13.8 &$\dots$	       &13.0  &$\dots$	      \\[3mm]
NGC\,5921   &$65_{-8}^{+4}$	  & $8.0_{-1.0}^{+0.5}$    &6.3  &$1.3_{-0.2}^{+0.1}$    &11.0  &$0.7_{-0.1}^{+0.1}$  \\[3mm]
NGC\,5964   &$55_{-13}^{+38}$   & $7.0_{-1.5}^{+4.5}$    &7.6  &$0.9_{-0.2}^{+0.6}$    &4.3   &$1.6_{-0.3}^{+1.0}$  \\[3mm]
NGC\,6946   &$254_{-63}^{+?}$   & $8.0_{-2.0}^{+?}$      &7.6  &$1.1_{-0.3}^{+?}$      &7.0   &$1.1_{-0.3}^{+?}$    \\[3mm]
NGC\,7479   &$94_{-25}^{+6}$    & $13.0_{-4.0}^{+1.0}$   &11.8 &$1.1_{-0.3}^{+0.1}$    &10.0  &$1.3_{-0.4}^{+0.1}$  \\[3mm]
NGC\,7741   &$109_{-17}^{+?}$   & $5.5_{-1.0}^{+?}$      &5.6  &$1.0_{-0.2}^{+?}$      &7.5   &$0.7_{-0.2}^{+?}$    \\[3mm] \hline 
\end{tabular}
\label{tab:CRoverLength}
\end{table*}

\begin{table*}
\caption{The pattern speed \Op, $r(CR)$, and ILR radius for our sample galaxies.}
\centering \tabcolsep=8pt
\begin{tabular}{llcclccl}
\hline \hline
 Object   &\Op	    	    & $r(CR)$	     & Resonance radius&\Op	        & $r(CR)$		& Resonance radius   & Structure    \\  
    \     &\kmsarc  	    & \arcsec	     & \arcsec    &\kmskpc	        &  kpc		& kpc	     &	   \\ \hline \\	        
IC\,342   & $0.59_{-0.09}^{+0.09}$& $344_{-79}^{+26}$  & \dots      &   $31_{-5}^{+5} $ & $6.5_{-1.5}^{+0.5}$ & \dots	     &Bar	   \\[2mm]  
NGC\,2403 & $0.34_{-0.02}^{+0.14}$& $390_{-32}^{+?}$   & \dots      &   $22_{-1}^{+9} $ & $6.1_{-0.5}^{+?}$	& \dots	     &Bar	   \\[2mm] 
NGC\,4294 & $2.92_{-0.81}^{+0.14}$& $29_{-6}^{+4}$     & \dots      &   $43_{-12}^{+3}$ & $1.9_{-0.4}^{+0.3}$ & \dots	     &Bar	   \\[2mm] 
NGC\,4519 & $1.46_{-0.58}^{+0.44}$& $>68$	     & 20 (ILR)   &   $20_{-8}^{+6} $ & $>5$		& 1.5 (ILR)    &Spiral	\\[2mm] 
NGC\,5371 & $2.57_{-0.18}^{+0.92}$& $93_{-16}^{+?}$    & \dots      &   $14_{-1}^{+5} $ & $17.0_{-3.0}^{+?}$  & \dots	     &Spiral 	\\[2mm] 
NGC\,5921 & $1.59_{-0.24}^{+0.24}$& $65_{-8}^{+4}$     & \dots      &   $13_{-2}^{+2} $ & $8.0_{-1.0}^{+0.5}$ & \dots	     &Spiral/Bar	\\[2mm] 
NGC\,5964 & $3.00_{-0.60}^{+0.12}$& $55_{-13}^{+38}$   & \dots      &   $25_{-5}^{+1} $ & $7.0_{-1.5}^{+4.5}$ & \dots	     &Spiral/Bar	\\[2mm] 
NGC\,6946 & $0.79_{-0.19}^{+0.19}$& $254_{-63}^{+?}$   & 54 (OILR)  &   $25_{-6}^{+6} $ & $8.0_{-2.0}^{+?}$	& 1.7(OILR)    &Spiral/Oval	\\[2mm] 
NGC\,7479 & $2.83_{-0.47}^{+0.47}$& $94_{-25}^{+6}$    & \dots      &   $18_{-3}^{+3} $ & $13.0_{-4.0}^{+1.0}$& \dots	     &Bar	   \\[2mm] 
NGC\,7741 & $1.13_{-0.36}^{+0.48}$& $109_{-17}^{+?}$   & 22 (ILR)   &   $19_{-6}^{+8} $ & $5.5_{-1.0}^{+?}$	& 1.3 (ILR)    &Bar	   \\[2mm] \hline
\end{tabular}\\
{In the rightmost column, we specify the morphological structure to which the pattern speed has been attributed.}
\label{tab:resonances}
\end{table*}

\subsection{Fourier Decomposition}
\label{sec:m2}
As an alternative to analysing the ellipticity profiles, the isophotal shapes can be described by means of the Fourier decomposition method \citep{ElmegreenElmegreen1985ApJ, Buta1986I, Buta1986II, Aguerrietal1998}. Certain preference can be attributed to this method as it avoids confusion caused by the morphological changes that accompany the onset of spiral structure in the disk, which could artificially lengthen the bar. Simulations predict that Fourier decomposition of the azimuthal luminosity profile is dominated by the ($m=2$) at the bar and falls sharply when the bar ends \citep{Ohtaetal1990}, however, in reality, matters are more complicated, and a combination of the Fourier components or the phase shift of the second component can be used to derive the length of a bar \citep{AthanassoulaMisiriotis2002,Aguerrietal2000,Aguerrietal2003,ZhangButa2007}.

\begin{figure*}
\resizebox{\hsize}{!}{\includegraphics{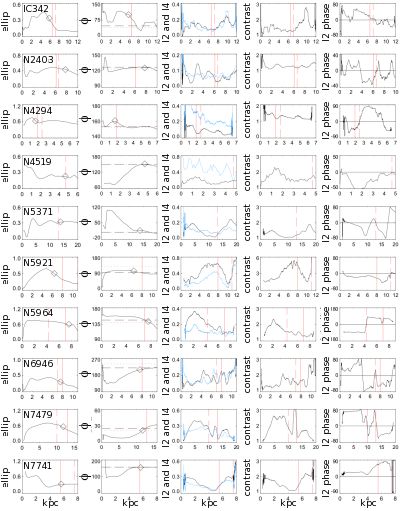}}
\caption{From left to right: Projected ellipticity and position angle $\phi$ profiles for our sample galaxies, deprojected $2^{nd}$ (black) and $4^{th}$ (blue) Fourier component profiles, contrast parameter, and phase shift of the second component. In each panel, the vertical solid line indicates $r(CR)$, the diamond, bar radius derived from the ellipticity profile $r(bar)_\epsilon$, and the vertical dashed line, the bar radius from the Fourier decomposition $r(bar)_F$. In the $\phi$ column, the horizontal dashed line shows the outer disk $\phi$ from the near infrared image, and the dotted (red) line shows the \pap\ used when applying the \tw\ method.}
\label{fig:morphology}
\end{figure*}

We apply the Fourier decomposition to the deprojected Spitzer, William Herschel, and Calar Alto near infrared images described in section~\ref{sec:ellipticities}, and present the second and forth components ($I_2, I_4$) in Fig.~\ref{fig:morphology} \citep[see ][ for a detailed presentation of the method]{ElmegreenElmegreen1985ApJ}. In this paper, since we are mainly interested in the lengths of the bars in our sample, we do not analyse the Fourier decomposition extensively, but only use it to ensure that these profiles indicate the presence of a bar in agreement with those derived from the ellipse fitting analysis. We use a combination of the following criteria to determine the bar lengths ($r(bar)_F$), and where these three criteria agree, we can specify the end of the bar quite well.
\begin{enumerate}
\item The bar ends after the $m=2$ component has been dominant, and starts decreasing accompanied by an increase of the odd components. This indicates that the two-fold symmetry of the bar component has decreased.
\item The contrast parameter defined by \citet{Ohtaetal1990}, $I_b/I_{ib} \; = \; (1+I_2+I_4+I_6) \, / \, (1-I_2+I_4-I_6)$ is a measure of the intensity of the bar divided by the intensity of the inter-bar. As these authors have demonstrated, the contrast parameter is smaller in the region of galaxies where the bulge component dominates, then increases and reaches its peak at the middle of the bar, and falls at the end of the bar. 
\item The phase shift of the second Fourier component ($m_2$) changes sign \citep{ZhangButa2007}. 
\end{enumerate}

We note in Fig.~\ref{fig:morphology}, that not all these three criteria are fulfilled in our profiles, most likely due to presence of strong star formation knots, foreground stars, and dust lanes. However, we note that in some cases, in the proximity of  $r(bar)_\epsilon$, the first two criteria could hold, and in all cases, the phase of the second Fourier component crosses the zero line. We thus define the phase shift crossing radius as the $r(bar)_F$. Table~\ref{tab:CRoverLength} shows that this method yields results similar to the ellipticity profile analysis outlined in section~\ref{sec:ellipticities}, and overall we derive comparable  $r(CR) / r(bar)$ ratios using either of the photometric analysis methods. One prominent exception is the galaxy, NGC\,4294, where the Fourier decomposition method unveils an extended 4-6 kpc oval structure which is not seen in ellipticity profile analysis or directly in the images. This oval structure could cause the severe under-estimation of our $r(CR) / r(bar)_F$ for this galaxy.

\subsection{Comparison With Simulations}
\label{sec:numsim}
Numerical simulations of barred galaxies predict the slowdown of bars due to the exchange of angular momentum between the dark matter halo and the galactic disk \citep[e.g., ][]{DebattistaSellwood1998, Athanassoula2003, Martinez-Valpuestaetal2006, SellwoodDebattista2006, WeinbergKatz2007}. Bars in the fast rotator regime are expected to have $r(CR)/r(bar)$ between 1.0 and 1.4, except for simulations of \citet{ValenzuelaKlypin2003} and \citet{TiretCombes2007,TiretCombes2008}, which obtained slow rotators  with $r(CR)/r(bar)>1.4$. This ratio strongly relies on the measurement of the bar radius, which is far from trivial to derive homogeneously from observations, however, the consistency check presented in section~\ref{sec:morphology} demonstrates that our morphologically derived $r(CR)/r(bar)$ ratios are consistent with those predicted by numerical simulations. The average value for our measurements is $r(CR)/r(bar) = 1.1$. As our main goal is to derive the \Op, the $r(CR)/r(bar)$ can serve as a useful inspection since its implication is that the bars in our sample late type spiral galaxies are fast bars, as predicted by models \citep[e.g., ][]{Laineetal1998, DebattistaSellwood2000, RSL2008}.

\section{Kinematic Results and Discussion}
\label{sec:results}
\subsection{Velocity Dispersion in Disks of Late-type Spiral Galaxies}
\label{sec:veldisp}
The galactocentric $\sigma$ profiles, illustrated in Fig.~\ref{fig:allmaps}, show that for most galaxies $\sigma$ is almost constant throughout the observed field with some general hint of decline toward the 25th magnitude radius. NGC\,5371 and NGC\,7479 both show smaller \Ha\ $\sigma$ values in the central $\approx 10$\arcsec\ with the outer parts following the same trend as the rest of the sample. NGC\,5921 has a deficit of ionised gas in the bar region, outside which it displays a very flat $\sigma$-profile. Consistent with the statistical results of \citet{ErwinSparke2002} for early-type bars and assuming that epicyclic approximation can be used here, we find that three of our 10 late-type spirals (NGC\,4519, NGC\,6946, NGC\,7741) show evidence for an inner Lindblad Resonance ILR (see section~\ref{sec:notes} and Fig.~\ref{fig:allmaps}). In all three cases, the circular component of the observed \Vlos\ (see section~3) shows a clear presence of a more rapidly rotating component within the ILR. Inside the ILR, the $\sigma$  values are generally higher than outside this region, with a particularly steep rise toward the nucleus, and as displayed in the rotation curves, the higher $\sigma$ values appear to accompany more rapid rotation. This might be linked to the scenario where pseudo-bulges are caused by bar dissolution \citep[see the detailed review by ][]{KormendyKennicutt2004}.

The results for our sample compare well with two-dimensional kinematic maps from \citet{FalconBarrosoetal2006}, \citet{Peletieretal2007}, and \citet{Bokeretal2008}. To first order, our spatial resolution allows us to distinguish between local effects such as winds from star-forming regions and supernovae, and global secular evolution effects. We find that the $\sigma$ values are higher at the location of the star-forming \HII\ regions \citep[c.f.,][]{Allardetal2006}, as compared with regions where no strong star formation occurs. This result is similar to that which we found in Paper~I, i.e. that locally outflows from \HII\ regions lead to a higher measured velocity dispersion close to the regions, while globally there is very little radial dependence of sigma, implying large scale uniformity throughout the disk in the summed inputs of the energy sources to the gas.

\subsection{Notes on Individual Galaxies}
\label{sec:notes}

\noindent
{\bf IC\,342} has angular size of more than 20 arcminutes and has a weak primary bar with $r(CR) \approx 4.7$\arcmin, in approximately north-south direction \citep{Crosthwaite2002}. IC\,342 contains extremely strong molecular gas in the innermost parts of the bar, where the presence of an ILR ($\approx 2$\arcsec) has been suggested \citep{Shethetal2002, TurnerHurt1992, Bokeretal1997}. At this radius a star-forming ring-like structure has been observed to include a $\approx 15$ Myr old supergiant-dominated central stellar cluster \citep{Bokeretal1997}. The near infrared spectra show no evidence for multiple stellar content within the bar, in support of this system being young. Although the exact molecular content in the bar has not been established, the \CO\ observations by \citet{TurnerHurt1992} and \citet{Sakamotoetal1999} suggest gas inflow due to the bar feeding the circumnuclear starburst. Inside the ILR ring, a deficit of molecular gas content has been observed, and our observations show a very weak \Ha\ signal. We find an overall patchy ionised gas distribution with a steeply rising rotation curve, fully consistent with both \CO\ and \HI\ rotation curves which stay around 200 \kms\ out to 15\arcmin\ \citep{Rogstadetal1973, YoungScoville1982, Sofue1996, Crosthwaiteetal2000}.

The bar \Op\ has been estimated to 40 \kmskpc\ by \citet{TurnerHurt1992}. Radial streaming motions above 5 \kms\ along  one arm were detected by \citet{Newton1980} who also found $r(CR)$ between 16 and 21 kpc, which they found consistent with a spiral density wave with \Op\ = 10 \kmskpc. \citet{Crosthwaiteetal2000} derived the bar \Op\ = 21 \kmskpc\ based on morphology of the spiral structure and $r(CR)$ at around 400\arcsec\ with no sign of an ILR. Applying the \tw\ method on the \Ha\ emitting ionised gas, we derive $\Omega_p = 31_{-5}^{+5}$ \kmskpc\ and $r(CR)\approx 6.5$ kpc. We cannot verify the presence of the location of the ILR due to the large spatial bins which we use in order to increase the signal in the central regions (see Fig.~\ref{fig:allmaps}). Finally, we confirm that the  \ghafas\ observations deliver a $\langle V \rangle$/$\langle X \rangle$ value comparable to that when using the inner $\approx 2$\arcmin\ section of the \fantomm\ data.\\

\noindent
{\bf NGC\,2403} is a high surface brightness member of the M~81 group with flocculent arms and bright \HII\ regions and a total star formation rate $\approx 0.1$ \Myr\ \citep{Kennicutt1989}. In the inner kpc, this galaxy contains about half as much molecular gas as its atomic hydrogen content \citep[e.g., ][]{ThornleyWilson1995}, suggesting that the formation of molecular hydrogen is constrained by the low surface density of the gas \citep{Helferetal2003}. The central gas surface densities are below the threshold star formation density of \citet{Kennicutt1989}. \citet{Fraternalietal2001} have detected anomalous neutral hydrogen extending beyond the disk plane of the galaxy interpreted as galactic fountains powered by stellar winds from massive stars and supernova explosions, which in turn could be triggered by the bar. Confirmed by a large sample of \HII\ regions, there is a gradient in O/H across NGC\,2403 with a slope of 0.1 dex kpc$^{-1}$ \citep{Garnettetal1997}, supporting the study by \citet{Beckmanetal1987} who found that non-circular motions are not very efficient in mixing the ISM in this galaxy. Although \citet{Schoenmakersetal1997} found, using \HI\ kinematics, that NGC\,2403 is axisymmetric to a high degree, we find that the ellipse fits and Fourier analysis of the near infrared images reveal the presence of a $\approx 6$ kpc bar ( see Fig.~\ref{fig:morphology}).

Our observed \Ha\ rotation curve shows a moderate increase to $\approx 60$ \kms\ in the central kpc followed by almost a solid-body type increase out to around 10 kpc radius. The curve is consistent with the \HI\ curve presented in \citet{Shostak1973}; however, due to strong feed-back from the numerous \HII\ regions in the central $\approx 200$ pc radius (confirmed by the large $\sigma$ values), we cannot constrain the rotation curve within this region. \citet{Shostak1973} derived the spiral pattern speed 10.5 \kmskpc, whereas, using the \tw\ method, we derive the bar pattern speed $\Omega_p = 22_{-1}^{+6}$ \kmskpc, and $r(CR)$ at 6 kpc, with no sign of Lindblad resonances (Fig.~\ref{fig:allmaps}).\\

\noindent
{\bf NGC\,4294} appears to interact with NGC\,4299 with a projected separation of 27 kpc. Both galaxies are similar in morphology, mass, luminosity, gas content, and both exhibit a high global star formation rate \citep[e.g., ][]{KoopmannKenney2004}. Moreover, \citet{Chungetal2007} found that due to its position in the Virgo cluster (0.7 Mpc from M~87) interaction with the intra-cluster medium could be plausible. NGC\,4294 hosts many \HII\ regions, a quiescent nucleus, and two rather faint flocculent spiral arms. The outer regions of the spiral structure are consistent with enhanced star formation at the radius at which the gas surface density in the disk becomes super-critical for star formation \citep[][]{ThornleyWilson1995, Thornley1996}. 

The observed \Ha\ velocities show streaming motions along the spiral arms \citep[][ and this work]{Cheminetal2006}. Although the receding part of the disk appears slightly perturbed, the overall appearance of the velocity field is quite regular. The \Ha\ rotation curve is consistent with that presented by \citet{Rubinetal1999}. Using the \tw\ method, we derive  $\Omega_p = 43_{-12}^{+3}$ \kmskpc, with $r(CR) \approx 1.9$ kpc in agreement with the location where the spiral arms emerge from the ends of the bar. Within the $r(CR)$, the velocity dispersion follows a plateau, whereas outside this radius it decreases steadily.\\

\noindent
{\bf NGC\,4519} hosts many \HII\ regions in the bar as well as the two partly broken spiral arms. The central regions exhibit significant molecular gas content \citep{Bokeretal2002}, although the star formation efficiency is constant in the inner two effective radii \citep{RowndYoung1999}. Our velocity field  displays a ``S''-shaped zero-velocity curve over the central 4 kpc radius region of the galaxy. The rotation curve rises steeply in the central few arcseconds, which indicates a rapidly rotating component within $\approx 30$\arcsec, and it continues to increase to 130 \kms\ at 6 kpc radius. This is consistent with the optical rotation curve from \citet{Rubinetal1999} and the \HI\ position-velocity diagram of \citet{Helouetal1984}.

We derive $\Omega_p = 20_{-8}^{+6}$ \kmskpc, which we assign to outer spiral arms. We use this \Op\ to locate an outer ILR radius at about 1.5 kpc (20\arcsec) and an inner ILR radius at around 200 pc. The outer ILR radius is, to within the errors, consistent with the \Ha\ disk scale length derived by \citet{Koopmannetal2006}.

Similar to \citet{Fathietal2007TW}, we apply the \tw\ method on the pixels interior to the ILR, and find that the inner bar is decoupled from the outer oval, as its pattern speed is $\Omega_p = 45$ \kmskpc. This \Op\ yields that the corotation of the inner bar is located at ILR of the outer oval. The method we have applied here enables us to determine the secondary \Op\ only with an uncertainty of 50\%, since to zero-th order, the \tw\ method applied to the central region has to assume that the outer bar or spiral arms are stationary. This is not a realistic scenario, so we use this higher pattern speed for the inner bar only as an indication that we can detect the phenomenon, and that the pattern speed is higher than that of the outer bar.\\

\noindent
{\bf NGC\,5371} is a grand design, possibly post starburst, spiral galaxy with a LINER nucleus \citep{Rushetal1993, Koornneef1993, Elfhagetal1996}. It hosts a prominent 2 kpc stellar bar $\approx 45$ degrees from the orientation of the outer disk major-axis with no significant change when comparing $B$ and $H$-band images \citep{Eskridgeetal2002}, and displays a star formation rate of $0.9$ \Myr\ \citep{GonzalezDelgado1997}. The rotation curve was measured by \citet{ZasovSilchenko1987} from which they derived an exponential main disk scale length of $\approx 8.7$ kpc, and a massive bulge ($M_{bulge} \geq M_{disk}$). Although the outer parts of their rotation curve are not well determined, the inner regions agree well with our \Ha\ rotation curve. 

Our observations display a deficiency of \Ha\ emission across the bar, in agreement with the \Ha\ images from \citet{GonzalezDelgadoPerez1997}. The velocity field displays clear disk-like rotation, with various ``wiggles'' in the zero-velocity curve, often interpreted as streaming motions along spiral arms \citep{Emsellemetal2006}. The rotation curve rises steadily in the central 30\arcsec, and reaches 250 \kms\ at 15 kpc radius. We derive $\Omega_p= 14_{-1}^{+5}$ \kmskpc\ and the $r(CR) \approx 17$ kpc, i.e., 93\arcsec. The deficiency of the \Ha\ emission from the stellar bar, combined with its non-optimal orientation compared with the outer disk major-axis, indicate that we are probing the \Op\ of the spiral arms, and not the well-known 2 kpc bar. 

We note that by placing corotation at 17~kpc, we bring the inner 4:1 resonance close to the end of the symmetric part of the spiral structure. At this region, in agreement with the predictions by \citet{Patsisetal1997}, a bifurcation of the northern arm is seen in Fig.~\ref{fig:allmaps}. Furthermore, the association of the end of the symmetric part of the spirals with the inner 4:1 resonance is supported by the radial variation of the ratio of the amplitudes of the $m=4$ to the $m=2$ components \citep{Grosbol1985, Patsis1991}.\\

\noindent
{\bf NGC\,5921} has a very poor molecular and neutral gas content \citep{Verter1985} and relatively poor ionised gas in a $\approx 8-12$ kpc prominent stellar bar which is enclosed by sharp inner ring-like structure. The bar contains two straight dust lanes, and the spiral arms host many \HII\ regions \citep{GonzalezDelgadoPerez1997, Aguerrietal1998, Rautiainenetal2005}, and form the ring at the bar ends, just before they become very open, broad, and diffuse. At the north end of the bar, the arm becomes brighter as it leaves the ring \citep{Eskridgeetal2002}. \citet{MartinFriedli1997} found that the stellar bar has an axis ratio of 0.5 and is oriented $18$ degrees in the north-east direction. These authors also derived an asymmetric star formation efficiency $<0.2$ \Myr\ in the bar, and seven times higher value in the circumnuclear region \citep{GonzalezDelgado1997}.

Like in NGC\,5371, the \Ha\ emission is so weak in the bar of this galaxy that we have to make use of large spatial bins to measure any kinematic information. The rotation curve cannot be constrained within the bar. The \tw\ method thus delivers the pattern speed of the prominent spiral structure ($\Omega_p= 13_{-2}^{+2}$ \kmskpc), with the spiral pattern $r(CR)\approx 8$ kpc \citep[or 65\arcsec, in good agreement with the $r(CR)=71$\arcsec\ of][]{RSL2008}, which is just outside the 58\arcsec\ radius of the strong stellar bar \citep{ButaZhang2009}. This suggests that the bar shares the same pattern speed as the spiral arms \citep[e.g., ][]{Taggeretal1987}.\\

\noindent
{\bf NGC\,5964} hosts a very prominent and elongated bar surrounded by diffuse, extended, and fragmented spiral arms with no regular pattern \citep{ElmegreenElmegreen1987}. This flocculent galaxy is rich in neutral Hydrogen content ($M_\mathrm{HI}/M_\mathrm{Tot} = 82\%$) and hosts a patchy circumnuclear structure \citep{GiovanelliHaynes1983, Bokeretal2002}. The \Ha\ rotation curve displays almost a solid body rotation in the inner 10 kpc. Our analysis shows that $\Omega_p= 25_{-5}^{+1}$ \kmskpc\ is well outlined in Fig.~\ref{fig:allmaps} with the location of the $r(CR)$ at 6.5 kpc. 

NGC\,5964 displays the largest misalignment ($36$ degrees) between the \pap\ and \pak, which could be caused by the projection effect since this galaxy is near face-on. Such misplacement can also be seen as an indication of strong streaming along the spiral arms and prominent bar. In Fig.~\ref{fig:allmaps} we show that the $r(CR)$ agrees with the radius where the patchy spiral structure starts, indicating that the bar and spiral arms rotate at the same rate \citep[see ][]{Taggeretal1987}.

Theoretical work such as that by \citet{ElmegreenThomasson1993} have shown that flocculent spirals can form via swing amplification of star formation patches, and thus should have no pattern speed. Although their proposed mechanism explains the patchiness of these galaxies, it does not exclude the presence of density waves in the galaxies where swing amplification builds patchy spiral structure. Support for such a scenario comes from simulations by \citep{WadaNorman2001} as well as observational studies by \citet{ThornleyMundy1997, GrosbolPatsis1998, ElmegreenElmegreenLeitner2003}. Notably, in NGC\,5055, \citet{ThornleyMundy1997} could outline the spiral arms by a spline fit to the $K$-band image and show that the non-circular motions in their residual \HI\ velocity field follow these arms. These authors combined the photometry with kinematics to derive the pattern speed for the underlying density wave at 35-40 \kmskpc, and conclude that in some flocculent galaxies the underlying density wave is difficult to observe due to their overall lower gas surface density (e.g., that in NGC\,5055 is a factor of 6 lower than the gas density in M\,51). To summarise, our results for NGC\,5964 imply an underlying density wave with $\Omega_p= 25_{-5}^{+1}$ \kmskpc. \\

\noindent
{\bf NGC\,6946} is a grand-design barred spiral galaxy with three main gravitational distortions: a large oval with radius $R\approx4.5$\arcmin, a $R\approx 1$\arcmin\ primary stellar bar with ellipticity 0.15, and a $R\approx 8$\arcsec\ nuclear bar with ellipticity 0.4, which is almost perpendicular to the main bar. We have presented a detailed analysis of the resonance structure in this galaxy in \citet{Fathietal2007TW}, and confirm that the maps presented in this paper are fully consistent with our previous work. \\

\noindent
{\bf NGC\,7479} is an extensively studied starburst galaxy \citep[e.g., ][]{Hawardenetal1986, Aguerrietal2000, Laine2001} with a massive, $2.2\times10^{10}$ \Msun\ stellar bar with axis ratio of 0.25-0.4 \citep{Burbidgeetal1960, Martin1995}. Although there is no evidence of recent interaction, \citet{LaineHeller1999} have found many kinematic and morphological features consistent with a minor merger event for this galaxy, proposing the hypothesis that this galaxy is on its way to evolve toward an earlier Hubble type. Several studies have found the $r(CR)$ to be between 45-106\arcsec\  \citep[][]{ElmegreenElmegreen1985, Quillenetal1995, BeckmanCepa1990} and \Op\ to be between 27 and 100 \kmskpc\  \citep{Laineetal1998, delRioCepa1998}. The mean star formation efficiency in the bar, although asymmetric, is 0.4 \Myr\ and increases to 0.63 \Myr\ in the circumnuclear regions \citep{MartinFriedli1997}, indicating inflow along the bar.

Our \Ha\ rotation curve is consistent with that derived by \citet{Garridoetal2005} with a prominent and rapidly rotating component in the central 30\arcsec. This curve, as well as the angular frequency curve, cannot be determined in the central kpc region due to the weak \Ha\ emission, and hence we cannot confirm the presence of the ILR reported by \citet{Laineetal1998}. We find $\Omega_p = 18_{-3}^{+3}$ \kmskpc, and $r(CR) \approx 13$ kpc. Our $r(CR)$ is almost twice the $55\arcsec$ found by \citet{PuerariDottori1997}, the $r(CR)$ suggested by numerical simulations of \citet{Laineetal1998}, and the $58\arcsec$ found from potential-density phase-shift analysis carried out by \citet{ButaZhang2009}. Using the $r(CR)$ found by these authors, our angular frequency curve implies $\Omega_p \approx 23$ \kmskpc, i.e., slightly above the value presented here. This would also make the bar with $r(CR)/r(bar)_F \approx 0.9$, to within the errors, comparable with the value we present in Table~\ref{tab:CRoverLength}.\\

\noindent
{\bf NGC\,7741} has a prominent bar with axis ratio 0.36, two short and flocculent spiral arms including a significant amount of diffuse ionised gas and  many \HII\ regions \citep[e.g., ][]{Duvaletal1991, Blocketal2004, Laurikainenetal2004}. This bulge-less galaxy has a mean star formation efficiency of 0.1 \Myr\  \citep{MartinFriedli1997}, with very weak \CO\ content in the bar \citep{Braineetal1993}. \citet{Duvaletal1991} found strong ionised gas flows in the bar region and a ratio of 30\% between the bar mass and that of the inner disk inside the bar radius.

Our rotation curve agrees with the low-resolution curve of \citet{Duvaletal1985} and \citet{Garridoetal2002}. The \tw\ method results in $\Omega_p = 19_{-6}^{+8}$ \kmskpc\ for the main bar with $r(CR) \approx 6.5$ kpc, c.f., \Op=31 \kmskpc\ from \citet{Duvaletal1985}. We find the location of an outer ILR at $\approx 1.3$ kpc, inside which the rotation curve suggests the presence of a dynamically cold rapidly rotating component. We find also an inner ILR at $\approx 300$ pc, and no hint of a higher pattern speed inside the outer or the inner ILR. In NGC\,7741, like in NGC\,7479, the ILR radii given here are based on the assumption that epicyclic approximation can be applied, and it should be noted that in strong bars, such as in NGC\,7741, where epicyclic approximation breaks down, the actual resonances may be in different locations \citep[e.g., ][]{ReganTeuben2004}.\\

\section{Conclusions}
\label{sec:conclusions}
Here, we have shown that an intensity-velocity map in \Ha\ of the type obtained using a Fabry-Perot interferometer, can be successfully used to derive the gas kinematics of disk galaxies with coverage and precision sufficient to derive the \Op\ for those where star formation is well spread across the galaxy. Although this technique is not new \citep{Hernandezetal2005TW, Emsellemetal2006, Fathietal2007TW}, its precision and the number of galaxies analysed here give our results considerable value. It is clear that the \Ha\ emitting gas does not act as a linear probe for the galaxy's surface density. However, only a fraction of the pixels in our maps are attributed to the compact \HII\ regions. The \tw\ method applied in this way gives us data points with high spatial and velocity resolution over a major fraction of the galaxy disk, and our results demonstrate that, with caution, the \tw\ method can be applied to deliver reliable pattern speeds in disk galaxies.

Although very simple, the radial tests that we have described in section~\ref{sec:radius} have proven useful to get a first order estimation of a possible secondary pattern speed. Fig.~\ref{fig:radialOpvalues} shows that the pattern speeds decline steadily as larger sections of the field are covered, and exactly how much of the disk needs to be covered to get reliable values depends very strongly on the size of the bars and/or spiral arms. 

Assuming that the epicyclic approximation can be used, the pattern speeds can be used to identify resonances, such as the ILR, in our sample galaxies. The presence of an ILR does not alone imply a decoupling of the region inside it, however, an ILR is necessary for a decoupling of nested bars \citep[e.g.,][]{EnglmaierShlosman2004}. We do trust the presence of secondary pattern speeds based on a combination of the radial behaviour of the $\langle V\rangle / \langle x\rangle/\sin(i)$, the presence of the ILR, and the presence of morphologically distinct features such as an inner bar. In galaxies without an ILR, we do not claim the presence of a secondary pattern speed, thus, once certain of the mere presence of a secondary pattern speed, its presence can be confirmed using the full slit coverage but only for slits falling in the region corresponding to the extent of the inner bar. The numeric value of the secondary pattern speed, on the other side, is far more uncertain.

We confirm our previous results from \citet{Fathietal2007TW} and since the uncertainties in deriving the secondary pattern speeds are as large as 50\%, we do not include the values here, however, we note that the detected secondary patterns are of the order of two to three times higher than the main pattern speeds. Similar demonstrations have been presented in previous studies by \citet{Corsinietal2003, Hernandezetal2005TW, Emsellemetal2006}, and \citet{Fathietal2007TW}. Overall the results from applying the \tw\ method on the \Ha\ maps of 10 late-type spiral galaxies can be listed as follows:

\begin{itemize}
\item By sectioning the observed velocity fields in galactocentric rings, we have been able to illustrate that $\langle V \rangle$/$\langle X \rangle$ is always highest at the centre, with a radially asymptotic decrease as pixels from larger sections of the disks are used, and converging towards the \Op\ when the full disk is covered.

\item Using the red continuum as our weighting function, we have found that the derived pattern speeds do not differ from those obtained when the \Ha\ surface brightness maps were used.

\item The \tw\ method applied along the photometric major-axis gives least scatter fits as compared with extracting the parameters along the kinematic major-axis. This further supports the presence of strong non-circular motions in barred spiral galaxies which could artificially induce misalignment between the photometric and kinematic position angles. Such non-circular motions are capable of transporting gas from the outer parts of the disks towards the circumnuclear regions, where secondary bars or inner disks can form.

\item Comparing the kinematically derived corotation radii for the bars with the bar radii independently derived from the morphology, we find that our $r(CR) / r(bar)$ are consistent with predictions from numerical models using static or live dark matter halos or modified gravity.

\end{itemize}

In six galaxies (IC\,342, NGC\,2403, NGC\,4294, NGC\,4519, NGC\,6946, NGC\,7479, and NGC\,7741), we have directly derived the \Op\ of the main bar, and in four galaxies (NGC\,4519, NGC\,5371, NGC\,5921, and NGC\,5964) we have been able to derive the \Op\ of the spiral arms. Dynamically, bars and spiral arms can interact directly when they are corotating \citep[e.g., NGC\,1365; ][]{Lindblad1999}, or by means of non-linear mode coupling in which case the $r(CR)$ of the bar and the ILR of the spiral arms overlap \citep[e.g., ][]{Taggeretal1987, MassetTagger1997}. In two galaxies (NGC\,5921 and NGC\,5964) we have found that the $r(CR)$ of the spiral structure roughly coincides with the $r(CR)$ of the bar, which suggests a rich variety of resonant interaction that have been predicted by theory, and our derived \Op s complement the values for 10 late-type barred spiral galaxies in excellent agreement with \Op s from numerical simulations using a live dark matter halo or modified gravity.

We use epicyclic approximation, as commonly used in models, to identify the location of the resonances in our sample galaxies. Three of our ten late-type spirals (NGC\,4519, NGC\,6946, NGC\,7741) show evidence for an ILR, two of which suggest the presence of a secondary pattern speed. In all three galaxies, the rotation curve shows a clear presence of a more rapidly rotating component within the ILR, suggesting that they harbour substantial amounts of the interstellar medium in the central regions. In all three cases, the \Ha\ velocity dispersion is higher inside the ILR with a steep rise toward the nucleus. The combination of these two observational features supports the idea that the thickening of the gaseous component in the central region could build a bulge-like component in late-type spirals (see also Paper~I). We note however that in strong bars such as in NGC\, 7741, epicyclic approximation could break down, and consequently, the location of resonances would not be valid.

The present paper is the second in a series in which we will pursue detailed dynamical analysis of late-type spiral galaxies. Our current results will be complemented with a full harmonic analysis of the velocity fields followed by detailed numerical simulations, and more detailed estimates of the systematic effects due to non-continuity in the \Ha\ emitting gas by applying the \tw\ method to high resolution simulations. Furthermore, a combination of these information with a robust estimation of the bar and spiral strengths could help us to better understand the evolution of structure in spiral galaxies.

\section*{Acknowledgments}
We thank the anonymous referee for insightful comments which helped us clarify some important points and revise and improve the quality of this work. We also thank Isabel P\'erez for fruitful discussions. This work is based in part on observations made with the Spitzer Space Telescope as well as the use of the NASA/IPAC Extra-galactic Database (NED) both operated by the Jet Propulsion Laboratory, California Institute of Technology, under contract with the National Aeronautics and Space Administration (NASA). We also acknowledge the extensive use of Google, the VizieR service for accessing online catalogues, as well as NASA's Astrophysics Data System Bibliographic Services. CC and OH acknowledge support from NSERC, Canada and FQRNT, Qu\'ebec. KF and JEB acknowledge support from the Spanish Ministry of Education and Science Project AYA2007, and the IAC project P3/86. KF acknowledges support from the Wenner-Gren foundations and the Swedish Research Council (Vetenskapsr\aa det), and the hospitality of the European Southern Observatory (ESO, Garching) where parts of this work were carried out.

\bibliographystyle{apj}
\bibliography{K_Fathi-TWpaper}

\end{document}